\documentclass[letter,12pt]{article}

\usepackage{pstricks}
\usepackage{epsfig}
\usepackage{amsmath}
\usepackage{bm}
\usepackage{array}
\usepackage{subfig}
\usepackage{graphicx}
\usepackage{cite}
\usepackage{amssymb}

\def\({\left(}
\def\){\right)}

\begin{document}
\setlength{\baselineskip}{18pt}
\vspace{-3cm}
\begin{flushright}
OSU-HEP-12-04\\
UMD-PP-012-004
\end{flushright}

\renewcommand{\thefootnote}{\fnsymbol{footnote}}

\begin{center}
{\Large\bf \boldmath{$B-L$} Violating Nucleon Decay and \\[0.1in]
GUT Scale Baryogenesis in \boldmath{$SO(10)$}}\\
\end{center}

\vspace{0.5cm}
\begin{center}
{ \bf {}~K.S. Babu}$^a$\footnote{Email:
babu@okstate.edu} and {\bf R.N. Mohapatra}$^b$\footnote{Email: rmohapat@umd.edu}
\vspace{0.5cm}

{\em $^a$Department of Physics, Oklahoma State University,
Stillwater, OK 74078, USA }
\vspace*{0.3cm}

{\em $^b$Maryland Center for Fundamental Physics, Department of Physics,\\ University of Maryland,
College Park, MD 20742, USA }

\end{center}

\begin{abstract}
\setlength{\baselineskip}{18pt}

We show that grand unified theories based on $SO(10)$ generate naturally the
next--to--leading baryon number violating operators of dimension seven.
These operators, which violate $(B-L)$, lead to unconventional decays of the
nucleon such as $n \rightarrow e^-K^+, e^- \pi^+$ and $p \rightarrow \nu \pi^+$.
In two--step breaking schemes of non-supersymmetric $SO(10)$, nucleon lifetime
for decays into these modes is found to be within reach of experiments.
We also identify supersymmetric scenarios where these decays may be accessible,
consistent with gauge coupling unification.
Further, we show that the $(B-L)$--asymmetry generated in the decays of GUT scale scalar
bosons and/or gauge bosons can explain consistently the observed baryon asymmetry of the universe.
The induced $(B-L)$ asymmetry is sphaleron--proof, and survives down to the weak scale
without being erased by the electroweak interactions. This mechanism works efficiently
in a large class of non--SUSY and SUSY $SO(10)$ models, with either a $126_H$ or a $16_H$ Higgs field employed
for rank reduction. In minimal models the induced baryon asymmetry is tightly connected to
the masses of quarks, leptons and neutrinos and is found to be compatible with observations.
\end{abstract}

\newpage
\renewcommand{\thefootnote}{\arabic{footnote}}
\setcounter{footnote}{0}

\section{Introduction}

Baryon number  violation is a very sensitive probe of physics beyond the Standard Model (SM).
Interactions which violate baryon number ($B$) are not present in the renormalizable part of the SM
Lagrangian, but they can arise as effective higher dimensional operators. The lowest $B$--violating
operators \cite{weinberg} have $d=6$ and are
suppressed by two powers of an inverse mass scale. These operators are realized naturally when SM is
embedded in a grand unified theory (GUT) such as $SU(5)$ and $SO(10)$ upon integrating out the
heavy vector gauge bosons and colored scalar bosons. They lead to the decay of the nucleon into modes such as $p \rightarrow e^+ \pi^0$ and
$p \rightarrow \overline{\nu} K^+$.  Present experimental limits on nucleon lifetime constrain the
masses of the mediators (vector gauge boson or scalar bosons) to be larger than about $10^{15}$
GeV, which is close to the unification scale determined from the approximate meeting of the three gauge
couplings when extrapolated to higher energies.

An interesting feature of the $d=6$ baryon number violating operators is that
they conserve baryon number minus lepton number ($B-L$) symmetry, leading to
the selection rule $\Delta(B-L) =0$ for nucleon decay\cite{weinberg}.
Thus, observation of any decays which violate $\Delta(B-L) =0$ rule would hint at new dynamics
different from those responsible for the $d=6$ operators.  Decay modes in this category
include $p \rightarrow \nu \pi^+$, $n\rightarrow e^- K^+, e^- \pi^+$ etc, which all obey
the selection rule $\Delta(B-L) = -2$.  Observation of these decay modes would thus furnish
evidence against the simple GUT picture with one step breaking to the SM.

In this paper we study the next--to--leading $d=7$ operators, which obey the selection rule
$\Delta (B-L) = -2$ for nucleon decay \cite{weinberg2}, and show that they arise naturally within $SO(10)$
grand unified theories. In non--supersymmetric $SO(10)$ models with an intermediate scale we find the
nucleon lifetime for decay modes such as $n \rightarrow e^- K^+, e^- \pi^+$
to be within reach of ongoing and proposed experiments. We also identify SUSY $SO(10)$
models where these decays may be within reach, consistent with gauge coupling unification.
While we focus mainly on renormalizable $SO(10)$ models with $126_H$ of Higgs bosons employed
for rank reduction, we show that our results also hold for models with $16_H$ used for this purpose.

The second main result of this paper is a mechanism for generating the baryon asymmetry of the universe at the GUT epoch.
The way it comes about is as follows. There are heavy scalar bosons and gauge bosons in $SO(10)$
theories which generate the $d=7$ operators. These particles have $(B-L)$--violating two--body
decays, which can generate the observed baryon asymmetry of the universe naturally, as we show here.
This would  not be possible in the case of $(B-L)$--preserving decays of GUT scale particles such as the ones in $SU(5)$.
Although grand unified theories were thought to be the natural stage for implementing the Sakharov's conditions for baryogenesis \cite{sakharov}
up until the mid-1980's \cite{book}, this idea was practically
abandoned after the realization that the sphalerons \cite{kuzmin}, which violate $B+L$ symmetry,
would erase any baryon asymmetry that obeyed the $\Delta(B-L)=0$ selection rule.
This is because the effective interactions generated by sphalerons, the non-perturbative configuration of the weak interactions, are in thermal equilibrium for temperatures in the range $10^2 ~\rm{GeV} \leq
T \leq 10^{12}$ GeV, and violate $(B+L)$ symmetry. However, if baryon asymmetry was generated by  $(B-L)$--violating
decays of GUT scale particles, they would be immune to sphaleron destruction.
We show that this mechanism of baryogenesis, which also induces the $d=7$ $B$--violating operators, is very efficient
and occurs quite generically in $SO(10)$ models.  In minimal
models there is a tight connection between the induced baryon asymmetry and the masses of
quarks, leptons and the neutrinos.  We also note that the minimal renormalizable versions of these models \cite{babu,aulakh} have been extremely successful in describing neutrino masses and mixings \cite{babu}, and in particular predicted relatively large value for the neutrino mixing angle $\theta_{13}$, which is consistent with recent results from Daya Bay, T2K, Double-Chooz and MINOS experiments \cite{theta13}. The results of the present paper show that these models can also explain the observed baryon asymmetry in a manner closely connected to the neutrino oscillation parameters.

This paper is organized as follows.  In Sec. 2 we discuss the $d=7$ $B$ and $(B-L)$--violating operators.  In Sec. 3 we show how these operators
arise in unified $SO(10)$ theories, both in the non-supersymmetric version and in the SUSY version. In Sec. 4 we address nucleon decay lifetime
for the $(B-L)$--violating modes in $SO(10)$ models with an intermediate scale.  Sec. 5 is devoted to GUT scale baryogenesis mechanism
tied to the $d=7$ operators.  Here we show the close connection between baryon asymmetry and fermion masses and mixings.
Finally, we conclude in Sec. 6.

\section{Baryon number violating \boldmath{$d=7$} operators}

We begin by recalling the leading baryon number violating operators in the
Standard Model which have $d=6$.  There are five such operators with baryon number $B=+1$ \cite{weinberg,abbott-wise}:
\begin{eqnarray}
{\cal O}_1 &=& (d^c u^c)^* (Q_i L_j) \epsilon_{ij},~~
{\cal O}_2 = (Q_i Q_j)(u^c e^c)^* \epsilon_{ij},~~
{\cal O}_3 = (Q_i Q_j) (Q_k L_l) \epsilon_{ij} \epsilon_{kl} \nonumber \\
{\cal O}_4 &=& (Q_i Q_j) (Q_k L_l) (\vec{\tau} \epsilon)_{ij}\cdot (\vec{\tau}\epsilon)_{kl},~~
{\cal O}_5 = (d^c u^c)^*(u^c e^c)^* ~.
\label{dim6}
\end{eqnarray}
Here we have not shown the color contractions (which is unique in each term
via $\epsilon_{\alpha \beta \gamma}$), and we have suppressed the flavor indices.
We have followed the standard notation for fermion fields with all fields being left--handed.  Thus $u^c$ stands for the left--handed
antiparticle of $u_R$.  The spinor indices are contracted via the charge conjugation
matrix between fields in parentheses. $i,j = 1,2$ are the $SU(2)_L$ indices.  The complex conjugate
operators of Eq. (\ref{dim6}) would of course carry $B=-1$.

An interesting feature of the $d=6$ baryon number violating operators on Eq. (\ref{dim6}) is that
they all carry lepton number $L=1$ along with $B=1$. Consequently these operators preserve $(B-L)$.  Thus
nucleon decay mediated by these operators would obey the selection rule $\Delta (B-L) = 0$.
The decays $p \rightarrow e^+ \pi^0, \overline{\nu} K^+$ and $n \rightarrow \overline{\nu} \pi^0,
e^+ K^-$ would be allowed by this selection rule, while decays such as $p \rightarrow \nu K^+$
and $n \rightarrow e^- K^+, e^- \pi^+$, which require $\Delta(B-L) = -2$ would be forbidden.  Grand unified
theories based on $SU(5)$ and $SO(10)$ gauge symmetries generate the operators of Eq. (\ref{dim6})
suppressed by two inverse powers of GUT scale masses.  With the GUT scale near $10^{15}$ GeV, as
suggested by the approximate unification of the three gauge couplings, these operators lead to nucleon lifetimes of
order $10^{32}-10^{36}$ years for $(B-L)$ conserving modes, which are in the range that is currently being
probed by experiments.  Discovery of nucleon decay into  $\Delta(B-L) = -2$ channels such as $n \rightarrow e^-K^+, e^-\pi^+$
would however suggest that the underlying dynamics is quite different from that of the $d=6$ effective operators of Eq. (\ref{dim6}).

As already noted in the introduction, while GUT scale particles can generate a baryon
asymmetry in their $B$--violating decays, it was realized that interactions of the electroweak
sphalerons would wash out any such  asymmetry that conserves $(B-L)$.  GUT scale baryogenesis
thus went out of fashion after the discovery of sphalerons.  This was also in part due to the leptogenesis mechanism
\cite{fukugita} discovered
soon thereafter, which can elegantly explain the observed baryon asymmetry with a connection
to the small neutrino masses induced via the seesaw mechanism\cite{seesaw}.

Now we turn to the next--to--leading $B$--violating operators beyond those of  Eq. (\ref{dim6}),
which are of dimension seven.
These operators are interesting in that they carry $(B-L) = \pm 2$ \cite{weinberg2}.  While they are suppressed
by one additional power of a heavy mass scale, they can naturally lead to sphaleron--proof baryogenesis, as we show here.  In several
instances we also find that these operators may lead to observable $(B-L)$ violating nucleon decay.

There are nine $d=7$ baryon number violation operators with $B= +1$ listed below \cite{weinberg2,weldon-zee}:
\begin{eqnarray}
{\cal \tilde{O}}_1 &=& (d^c u^c)^* (d^c L_i)^* H^*_j \epsilon_{ij},~~~~~~{\cal \tilde{O}}_2 = (d^c d^c)^* (u^c L_i)^* H^*_j \epsilon_{ij}, \nonumber \\
{\cal \tilde{O}}_3 &=& (Q_i Q_j)(d^c L_k)^*H^*_l \epsilon_{ij} \epsilon_{kl},~~
{\cal \tilde{O}}_4 = (Q_i Q_j) (d^c L_k)^*H^*_l (\vec{\tau} \epsilon)_{ij}\cdot (\vec{\tau}\epsilon)_{kl}, \nonumber \\
{\cal \tilde{O}}_5 &=& (Q_i e^c) (d^c d^c)^*H^*_i,~~~~~~~~~~
{\cal \tilde{O}}_6 = (d^c d^c)^*(d^c L_i)^* H_i, \nonumber \\
{\cal \tilde{O}}_7 &=& (d^c D_\mu d^c)^*(\overline{L}_i \gamma^\mu Q_i),~~~~~~{\cal \tilde{O}}_8 = (d^c D_\mu L_i)^*(\overline{d^c} \gamma^\mu Q_i), \nonumber \\
{\cal \tilde{O}}_9 &=& (d^c D_\mu d^c)^* (\overline{d^c} \gamma^\mu e^c)~.
\label{dim7}
\end{eqnarray}
We have used the same notation as in Eq. (\ref{dim6}).  Here $H$ is the Standard Model Higgs doublet transforming under
$SU(3)_C \times SU(2)_L \times U(1)_Y$ as $(1,2,+1/2)$.  $D_\mu$ stands for the covariant derivative with respect to the
$SU(3)_C \times SU(2)_L \times U(1)_Y$ gauge symmetry.  Note that ${\cal \tilde{O}}_2$, ${\cal \tilde{O}}_5$  and ${\cal \tilde{O}}_6$ must
be antisymmetric in the down--flavor indices.
Operators of the type $(\overline{d^c} D \hspace{-1.4ex}/\, Q_i)(d^c L_j) \epsilon_{ij}$ are not written, since they are related
to those listed in Eq. (\ref{dim7}) by the equations of motion.  All vector and tensor operators can be Fierz--transformed
into the set of operators in Eq. (\ref{dim7}).

Note that all operators of Eq. (\ref{dim7}) carry $B=1$ and $L=-1$, and thus $(B-L) = + 2$, with the complex conjugates
operators carrying $(B-L) = -2$.  It is these operators which can mediate nucleon decay of the type $n \rightarrow e^- K^+,
e^-\pi^+$ and $p \rightarrow \nu \pi^+$.  The higher dimensionality of these operators would suggest naively that the nucleon decay lifetime would
be much longer than the ones obtained from Eq. (\ref{dim6}).  However, as we show below, in unified theories based on $SO(10)$
with an intermediate
scale, these decays may be accessible to experiments.  Most interestingly, these operators can naturally generate baryon
asymmetry of the universe at the GUT scale, which is facilitated by the fact that the electroweak sphaleron interactions
do not wash out a  $(B-L)$ asymmetry generated at such a scale.

In the supersymmetric version of the standard model, baryon number violation can arise through operators in the superpotential analogous
to Eq. (\ref{dim7}).  These superpotential operators would have dimension six.
Holomorphicity of the superpotential would however constrain the allowed operators.   There is a single
operator of dimension six  given by the superpotential coupling
\begin{equation}
{\cal \tilde{O}}_1^{\rm ~SUSY} = d^c d^c u^c L_i (H_u)_j \epsilon_{ij},
\label{W}
\end{equation}
which carries $B=-1$ and $B-L = -2$.  This operator must be antisymmetric in the down-flavor indices owing to Bose symmetry.
$H_u$ here is the up--type Higgs doublet of MSSM.
In the superpotential, this operator will appear with two inverse powers of a heavy mass scale.
Terms in the Lagrangian resulting from Eq. (\ref{W}) would have two fermion fields, one Higgs field and two superpartner scalar fields, for example.
When the superpartner scalar fields are converted to standard model fermions by a gaugino loop, effective $d=7$ operators of Eq. (\ref{dim7}) would be
generated, suppressed by a factor $(M^2 M_{\rm SUSY})^{-1}$, rather than $M^{-3}$ that occurs
for Eq. (\ref{dim7}) without SUSY, where $M$ is the heavy mass scale.
Therefore, potentially these SUSY contributions can be more significant for nucleon decay.

We can now present the complete list of $(B-L) = -2$ effective operators through $d=7$ in the SM by adding to Eq. (\ref{dim7}) operators
with $B=0, L=2$.  These operators have been classified in Ref. \cite{leung}.  While not directly related to nucleon decay,
these operators arise along with the $d=7$ operators of Eq. (\ref{dim7}) in $SO(10)$ unified theories, and they are also
relevant for GUT scale baryogenesis.  Here we collect the linearly independent set of these operators through $d=7$.
The leading operator of course is the well-known $d=5$ seesaw operator \cite{weinberg}
\begin{equation}
{\cal O}_{d=5} = (L_i L_j) \, H_k H_l\, (\vec{\tau} \epsilon)_{ij} \cdot (\vec{\tau} \epsilon)_{kl}~.
\label{d5}
\end{equation}
The next--to--leading operators are of $d=7$, and there are ten of them, as listed below.
\begin{eqnarray}
{\cal O}_1' &=& (L_i L_j) (L_k e^c)\, H_l\, \epsilon_{ij} \epsilon_{kl},~~~~{\cal O}_2' = (L_i L_j) (L_k e^c) \,H_l\, (\vec{\tau} \epsilon)_{ij} \cdot
(\vec{\tau}\epsilon)_{kl}, \nonumber \\
{\cal O}_3' &=& (L_i L_j) (Q_k d^c)\, H_l \,\epsilon_{ij} \epsilon_{kl},~~~{\cal O}_4' = (L_i L_j) (Q_k d^c)\, H_l\, (\vec{\tau} \epsilon)_{ij} \cdot
(\vec{\tau}\epsilon)_{kl}, \nonumber \\
{\cal O}_5' &=& (L_i Q_j) (L_k d^c)\, H_l \, \epsilon_{ij} \epsilon_{kl},~~~{\cal O}_6' = (L_i Q_j) (L_k d^c) \, H_l\, (\vec{\tau} \epsilon)_{ij} \cdot
(\vec{\tau}\epsilon)_{kl}, \nonumber \\
{\cal O}_7' &=& (L_i L_j) (Q_k u^c)^*\, H_k\, \epsilon_{ij},~~~~{\cal O}_8' = (L_i L_j) (Q_k u^c)^* \,H_l\, (\vec{\tau}\epsilon)_{ij} \cdot
(\vec{\tau} \epsilon)_{kl}, \nonumber \\
{\cal O}_9' &=& (L_i d^c)(e^c u^c)^*\, H_j\, \epsilon_{ij},~~~~~{\cal O}_{10}' = (L_i D_\mu L_j)(\overline{u^c} \gamma^\mu d^c)\, \epsilon_{ij}~.
\label{lep}
\end{eqnarray}
We shall see the appearance of some of these operators in the embedding of Eq. (\ref{dim7}) in $SO(10)$ models.  In the supersymmetric
standard model, the four holomorphic operators ${\cal O}_1'-{\cal O}_4'$ would be allowed in the superpotential (with no significance attributed
to the spinor contractions of Eq. (\ref{lep}) when applied to the superfields).

\section{Origin of \boldmath{$d=7$} {\boldmath $B$}--violating operators in {\boldmath $SO(10)$}}

In this section, we show that the $d=7$ baryon number violating operators of Eq. (\ref{dim7}) arise naturally in the context of $SO(10)$ unified theories after the spontaneous breaking of $(B-L)$, which is a part of the gauge symmetry.
The $(B-L)$ symmetry may break at the GUT scale so that $SO(10)$ breaks directly
to the Standard Model gauge symmetry, or it may break at an intermediate scale $M_I$ below the GUT scale.  In the latter case the
intermediate symmetry could be one among several possibilities:
$SU(4)_C \times SU(2)_L \times SU(2)_R$; $SU(4)_C \times SU(2)_L \times U(1)_R$;
$SU(3)_C \times SU(2)_L \times SU(2)_R \times U(1)_{B-L}$; $SU(3)_C \times SU(2)_L \times U(1)_R \times U(1)_{B-L}$; or $SU(5) \times U(1)$,
with or without left--right parity symmetry.  In the non--supersymmetric version an intermediate scale is necessary to be
compatible with gauge coupling unification\cite{chang}, while with supersymmetry the direct breaking of $SO(10)$ down
to the MSSM is preferable.  Even in the latter case, there is room for intermediate scale particles,
provided that they form complete multiplets of the $SU(5)$ subgroup, since such particles
do not spoil the unification of gauge couplings observed with the MSSM spectrum.

To see how the $d=7$ operators of Eq. (\ref{dim7}) arise within $SO(10)$, we first focus on the scalar--mediated operators and  write down the Yukawa couplings in the most
general setup.  The Higgs fields which can couple to the fermion bi-linears $16_i 16_j$ are $10_H$, $\overline{126}_H$ and $120_H$,
with the couplings of the $10_H$ and $\overline{126}_H$ being symmetric in flavor indices $(i,j)$ and those of the $120_H$ being
antisymmetric.  The terms in these Yukawa couplings that are relevant to the generation of the $d=7$ operators are given below \cite{aulakh1,nath}.
\begin{eqnarray}
{\cal L}(16_i 16_j 10_H) &=& h_{ij} \left[ (u^c_i Q_j + \nu^c_i L_j)\, h - (d^c_i Q_j + e^c_i L_j)\, \overline{h} +
\left( \frac{\epsilon}{2} Q_i Q_j + u^c_i e^c_j - d^c_i \nu^c_j \right)\omega  \right. \nonumber \\
&& \left. + \left(\epsilon u^c_i d^c_j + Q_i L_j  \right)\omega^c \right],
\label{Yuk10}
\end{eqnarray}
\begin{eqnarray}
{\cal L}(16_i 16_j \overline{126}_H) &=& f_{ij} \left[(u^c_iQ_j - 3 \nu^c_i L_j)\,h - (d^c_iQ_j-3 e^c_iL_j)\,\overline{h} \right. \nonumber \\
&& \left.  + \sqrt{3}i\left(\frac{\epsilon}{2} Q_i Q_j - u^c_i e^c_j+ \nu^c_i d^c_j\right) \omega_1
+ \sqrt{3} i (Q_i L_j - \epsilon u^c_i d^c_j)\,\omega_1^c \right. \nonumber \\
&& \left.  + \sqrt{6}(d^c_i \nu^c_j + u^c_i e^c_j)\, \omega_2
+  2 \sqrt{3} i\, d^c_i\, L_j\, \rho  - 2 \sqrt{3} i\, \nu^c_i\, Q_j\, \overline{\rho} + 2 \sqrt{3} \, u^c_i \,\nu^c_j\,\eta \right. \nonumber \\
&& \left.- 2 \sqrt{3} i\, u_i^c \, L_j \, \chi + 2 \sqrt{3} i\,  e_i^c \,Q_j \,\overline{\chi} -2 \sqrt{3}\, d_i^c\, e_j^c\, \delta +
\sqrt{6} i \,Q_i\, L_j \,\overline{\Phi}
+ ....\right],
\label{Yuk126}
\end{eqnarray}
\begin{eqnarray}
{\cal L}(16_i 16_j 120_H) &=& g_{ij}\left[(d_i Q^j+e^c_iL_j)\,\overline{h}_1 - (u^c_i Q_j + \nu^c_i L_j)\, h_1 - \sqrt{2} Q_i L_j\, \omega_1^c \right. \nonumber \\
&& \left. - \sqrt{2} (u^c_i e^c_j - d^c_i \nu^c_j)\,\omega_1 -\frac{i}{\sqrt{3}}(d^c_i Q_j-3 e^c_iL_j)\, \overline{h}_2 + \frac{i}{\sqrt{3}}
(u^c_i Q_j-3 \nu^c_i L_j)\, h_2 \right. \nonumber \\
&& \left. -2 e^c_i Q_j \,\overline{\chi}+ 2 \nu^c_i Q_j\, \overline{\rho}-2 d^c_iL_j \, \rho + 2 u^c_i L_j \, \chi \right. \nonumber \\
&& \left. - i\, \epsilon\, d^c_i d^c_j \, \overline{\eta}+ 2 \,i\, u^c_i \nu^c_j\, \eta + \sqrt{2}\, i\, \epsilon\, d^c_i u^c_j \, \omega_2^c + \sqrt{2}\, i\,
(d^c_i \nu^c_j-e^c_i u^c_j)\, \omega_2 \right. \nonumber \\
&& \left. -\frac{\epsilon}{\sqrt{2}}Q_i Q_j \Phi - \sqrt{2}\, Q_i L_j \overline{\Phi}  -2 \,i\, d_i^c\,  e^c_j\, \delta + i\, \epsilon\,
 u_i^c\, u_j^c\, \overline{\delta} + ...\right].
\label{Yuk120}
\end{eqnarray}
These terms are written in terms of the Standard Model decomposition of the sub-multiplets.  We have followed the phase convention
of Ref. \cite{aulakh1}.  $\epsilon$ stands for the $SU(3)_C$ tensor $\epsilon_{\alpha \beta \gamma}$.
Here we have not displayed terms that are irrelevant for inducing the $d=7$ baryon number violating operators.  (Specifically, we
have omitted color singlet, color octet, and color sextet couplings.)  The Yukawa couplings obey $h_{ij} = h_{ji}$,
$f_{ij} = f_{ji}$ and $g_{ij}= - g_{ji}$.  The $SU(3)_C \times SU(2)_L \times U(1)_Y$ quantum numbers of the various sub-multiplets are given as
follows.
\begin{eqnarray}
&~& h(1,2,+1/2),~~~~\overline{h}(1,2,-1/2),~~~~ \omega(3,1,-1/3),~~~~\omega^c(\overline{3}, 1, 1/3),~~~~\nonumber \\
&~& \rho(3,2,1/6),~~~~ \overline{\rho}(\overline{3},2,-1/6),~~~~\eta(3,1,2/3),~~~~\overline{\eta}(\overline{3},1,-2/3),~~~~\nonumber \\
&~& \Phi(3,3,-1/3),~~~~\overline{\Phi}(\overline{3},3,1/3),~~~~\chi(3,2,7/6),~~~~\overline{\chi}(\overline{3},2,-7/6),~~\nonumber \\
&~& \delta(3,1,-4/3),~~~~\overline{\delta}(\overline{3},1,4/3)~.
\label{higgs}
\end{eqnarray}
Different fields with the same SM quantum numbers appear in some couplings, they are distinguished by subscripts $1,2$ etc.  We have
used the same notation for fields with the same SM quantum numbers in $10_H$, $\overline{126}_H$ and $120_H$, but it should
be understood that these are distinct fields.  After GUT symmetry breaking various subfields with the same SM quantum
number would mix.  Some of these mixings would involve the vacuum expectation value of the SM singlet field from the $\overline{126}_H$, denoted by $\Delta^c$ carrying $(B-L) = -2$.  It is this field that supplies large Majorana mass for
the right--handed neutrino through the coupling $f_{ij}\sqrt{6} \nu^c_i \nu^c_j \Delta^c$.  With $\left\langle \Delta^c \right\rangle
\neq 0$, trilinear scalar couplings of the type $\rho^* \omega H$, $\eta^* \rho H$, $\rho^* \Phi H$ and $\chi^* \eta H$, will develop.  This issue will be addressed in more detail below, but
we note that such couplings are invariant under the unbroken SM gauge symmetry.
When combined with the Yukawa couplings of Eqs. (\ref{Yuk10})-(\ref{Yuk120}), they would induce the $d=7$ baryon number violating operators of Eq. (\ref{dim7}).

\begin{figure}[h]
\centering
	\includegraphics[scale=0.5]{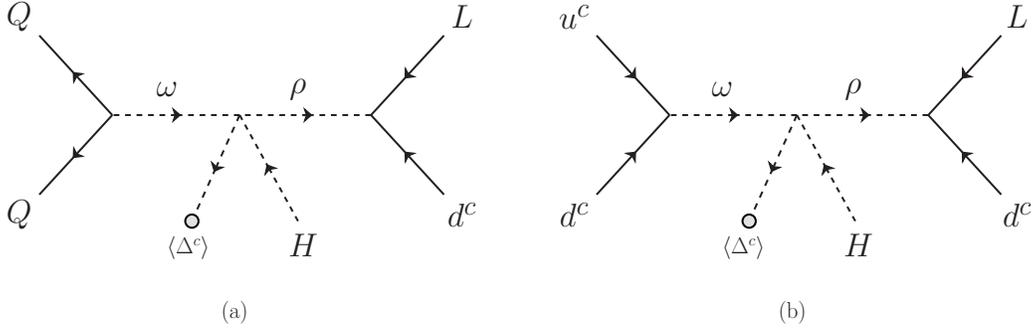}
	\caption{Effective baryon number violating $d=7$ operators induced by the symmetric Yukawa couplings of $10_H$ and $\overline{126}_H$ of $SO(10)$. Here the SM quantum numbers of the various fields are $\omega(3,1,-1/3),\,\rho(3,2,1/6)$, and $H(1,2,1/2)$. }
	\label{sym}
\end{figure}

To see how the $d=7$ operators arise in more detail, let us focus of the flavor symmetric Yukawa couplings of Eqs. (\ref{Yuk10})-(\ref{Yuk126}).
These couplings generate two of the $d=7$ operators as shown in Fig. \ref{sym}.  Here $H(1,2,1/2)$ is the SM Higgs doublet, which is a linear
combination of the $h(1,2,1/2)$ and $\overline{h}^*(1,2,-1/2)$ fields from $10_H$, $\overline{126}_H$ as well as any other Higgs sub-multiplet with the quantum number of $(1,2,1/2)$ in the theory with which these fields mix.  Similarly, $\omega(3,1,-1/3)$ generically stands for
any linear combination of $\omega$ and $(\omega^c)^*$ from the
$10_H$, $\omega_1, \omega_2$ and $(\omega_1^c)^*$ from the $\overline{126}_H$, etc.  Before estimating the strength of these operators,
let us examine the origin of the trilinear scalar couplings that appear in these diagrams in $SO(10)$.

To see the origin of $\rho^* \omega H$ and similar vertices, let us recall first the decomposition of various $SO(10)$ fields under the subgroups $SU(4)_C \times SU(2)_L \times SU(2)_R$ and $SU(5) \times U(1)$.
Under $G(2,2,4) \equiv SU(2)_L \times SU(2)_R \times SU(4)_C$, we have the following decomposition:
\begin{eqnarray}
16&=&(2,1,4)+(1,2,\overline{4}) \nonumber \\
10 &=& (2,2,1) + (1,1,6) \nonumber \\
126 &=& (1,1,6) + (3,1,10) + (1,3,\overline{10}) + (2,2,15) \nonumber \\
120 &=& (2,2,1) + (1,1,10)+(1,1,\overline{10})+(3,1,6)+(1,3,6)+(2,2,15)~.
\end{eqnarray}
Under  the $G(5,1) \equiv SU(5) \times U(1)$ subgroup various fields decompose as follows.
\begin{eqnarray}
16 &=& 1(-5) + \overline{5}(3) + 10(-1) \nonumber \\
10 &=& 5(2) + \overline{5}(-2) \nonumber \\
126 &=& 1(-10) + \overline{5}(-2) + 10(-6) + \overline{15}(6) + 45(2) + \overline{50}(-2) \nonumber \\
120 &=& 5(2) + \overline{5}(-2) + 10(-6) + \overline{10}(6) + 45(2) + \overline{45}(-2)
\end{eqnarray}

Now, the quartic coupling $(126)^4$, which is invariant (there is a single such coupling), contains the term $(2,2,15) \cdot (2,2,15) \cdot (1,1,6) \cdot (1,3,\overline{10})$ under $G(2,2,4)$.
The $\rho^*(\overline{3},2,-1/6)$ field is a subset of  $(2,2,15)$ fragment.
The $H(1,2,1/2)$ field is part of
$(2,2,15)$, while $\omega(3,1,-1/3) \subset (1,1,6)$
and one such field is also part of $(1,3,\overline{10})$. Thus one sees that the coupling
$(2,2,15) \cdot (2,2,15) \cdot (1,1,6) \cdot (1,3,\overline{10})$ would contain
the term $\rho^* \omega H \overline{\Delta^c}$, where $\overline{\Delta^c}$ denotes the SM singlet field
from $126_H$ that acquires a GUT scale VEV.  In the $G(5,1)$
decomposition, $\rho^*(\overline{3},2,-1/6) \subset  \overline{15}(6)$, $H(1,2,1/2)
\subset (45,2)$, and $\omega(3,1,-1/3) \subset (45,2)$.  Thus $(126)^4$ contains the
term $ \overline{15}(6) \cdot  (45,2) \cdot  (45,2) \cdot 1(-10)$, which has the piece
$\rho^* \omega H \overline{\Delta^c}$.

There are three non-trivial invariants of he type $(126)^2 \cdot (126^*)^2$.  These couplings
also contain the term  $(2,2,15) \cdot (2,2,15) \cdot (1,3,10) \cdot (1,1,6)$, as can be seen by examining the decomposition under $G(2,2,4)$
and separately under  $G(5,1)$.\footnote{The $(126)^2 \,(\overline{126})^2$  coupling would also contain a term $(2,2,15) \cdot (2,2,15) \cdot (1,3,10) \cdot (1,3, \overline{10})$ under $G(2,2,4)$,
which has a $\rho^* \omega H \overline{\Delta^c}$ term in it.  However, the $\omega$ appearing here is from the $(1,3,10)$, which is $\omega_2$
of Eq. (\ref{Yuk126}).  $\omega_2$ coupling by itself does not violate baryon number \cite{marshak}, as is evident from
Eq. (\ref{Yuk126}).}  In an analogous
fashion one sees that the coupling $(126)^2 \cdot (126^* 10)$ contains $(2,2,15) \cdot (2,2,15) \cdot (1,3,10) \cdot (1,1,6)$, with the $(1,1,6)$ arising from the $10$.  To complete this
discussion we also note that there are three invariants of the type $(120)^2 \cdot (126)^2$, which contain $(1,3,6) \cdot (1,3,10) \cdot (2,2,15) \cdot (2,2,15)$, with the $(1,3,6)$ and one $(2,2,15)$ taken from the $120$.
The $(1,3,6)$ fragment contains the $\eta(3,1,2/3)$ of
Eq. (\ref{higgs}), which would enter the $d=7$ operators arising by integrating out
the flavor antisymmetric $120$ fragments.  In particular, this term would induce the needed $\rho^* \eta H^*$ vertex.
The $(1,3,6) \cdot (1,3,10) \cdot (2,2,15) \cdot (2,2,15)$ term also contains $\chi^* \eta H$ vertex, with $\chi  \subset (2,2,15)$,
$\eta \subset (1,3,6)$ from the $120$.
The $(120)^2 \cdot (126)^2$ invariant also contains the term
$(3,1,6) \cdot (1,3,10) \cdot (2,2,15) \cdot (2,2,15)$, with the $(3,1,6)$ and one $(2,2,15)$ taken from the $120$.  This piece
generates the terms $\Phi^* \rho H^*$ vertex, with $\Phi^* \subset (3,1,6)$.
Finally, there are two invariants of the type
$(120)^2  \cdot 126 \cdot 126^*$, which also contain the term $(1,3,6) \cdot (1,3,10) \cdot (2,2,15) \cdot (2,2,15)$.

In order to complete $SO(10)$ symmetry breaking, additional Higgs fields such as a $45$, $54$ or a $210$ is needed.
The interactions of these fields with $126$ and $120$ can generate further trilinear and quadrilinear couplings.  Take
for example the case of a $54$ employed for completing the symmetry breaking.  The trilinear couplings $(126)^2\, 54$ and $(\overline{126})^2\, 54$
are then invariant.  Noting that under $SU(5) \times U(1)$ subgroup, $54 = 15(4) + \overline{15}(-4) + 24(0)$, we see that
the latter coupling would contain a term $1(10) \cdot 15(-6) \cdot \overline{15}(-4)$.  This has a piece $\Delta^c(1,1,0)_{\overline{126}} ~ \rho(3,2,1/6)_{\overline{126}} ~ \rho^*(\overline{3},2,-1/6)_{54}$, which would mix the two $\rho$ fields once $\left \langle \Delta^c \right \rangle \neq 0$ develops,
breaking $(B-L)$ by two units.  (Note that the $\rho_{\overline{126}}$ and $\rho_{54}$ carry different $(B-L)$ charges.)
The cubic coupling $(\overline{126})^2\, 54$ also contains the terms $5(2) \cdot 5(2) \cdot \overline{15}(-4)$ and
$\overline{45}(-2) \cdot \overline{45}(-2) \cdot 15(4)$ under $SU(5) \times U(1)$.  These terms contain the couplings $\omega(3,1,-1/3)_{\overline{126}} ~\, h(1,2,1/2)_{\overline{126}}~\,
\rho^*(\overline{3},2,-1/6)_{54}$ and $\omega^c_{\overline{126}}~\, \overline{h}(1,2,-1/2)_{\overline{126}} \,~ \rho(3,2,1/6)_{54}$ respectively.
These are the desired trilinear couplings for the generation of the $d=7$ operators.  In this case, the mixing of
$\rho^*(\overline{3},2,-1/6)_{54}$ with the $\rho(3,2,1/6)_{\overline{126}}$ is utilized in order to connect the $\rho$ field with
the fermion fields in Fig. \ref{sym}.  Similar results follow from the quartic
couplings $(126)^2 \, (54)^2$ and $(126 \cdot \overline{126})\, (54)^2$.

While trilinear scalar couplings of the type $\eta^* \rho H$ and $\overline{\Phi} \rho H^*$ do arise for the $\overline{126}_H$ sub-multiplets,
these couplings do not directly lead to baryon number violation.  The $\eta$ field from $\overline{126}_H$ has the coupling $\eta \,u^c\, \nu^c$, while
$\overline{\Phi}$ has the coupling $\overline{\Phi}\, Q\, L$ (see Eq. (\ref{Yuk126})).  The exchange of $\eta-\rho$ from $\overline{126}_H$ would
lead to an effective operator $(u^c \nu^c) (d^c L_i) H_j \epsilon_{ij}$, while that of $\Phi-\rho$ would generate the operator
$(Q_i L_j)(d^c L_k) H_l (\vec{\tau}\epsilon)_{ij} \cdot (\vec{\tau} \epsilon)_{kl}$, which is operator ${\cal O}_6'$ of Eq. (\ref{lep}).
The  $\eta^* \rho H$ and $\overline{\Phi} \rho H^*$ couplings of the $\overline{126}_H$ sub-multiplets would however be relevant for
GUT scale baryogenesis, since they also violate $(B-L)$ symmetry.

\begin{figure}
\centering
	\includegraphics[scale=0.55]{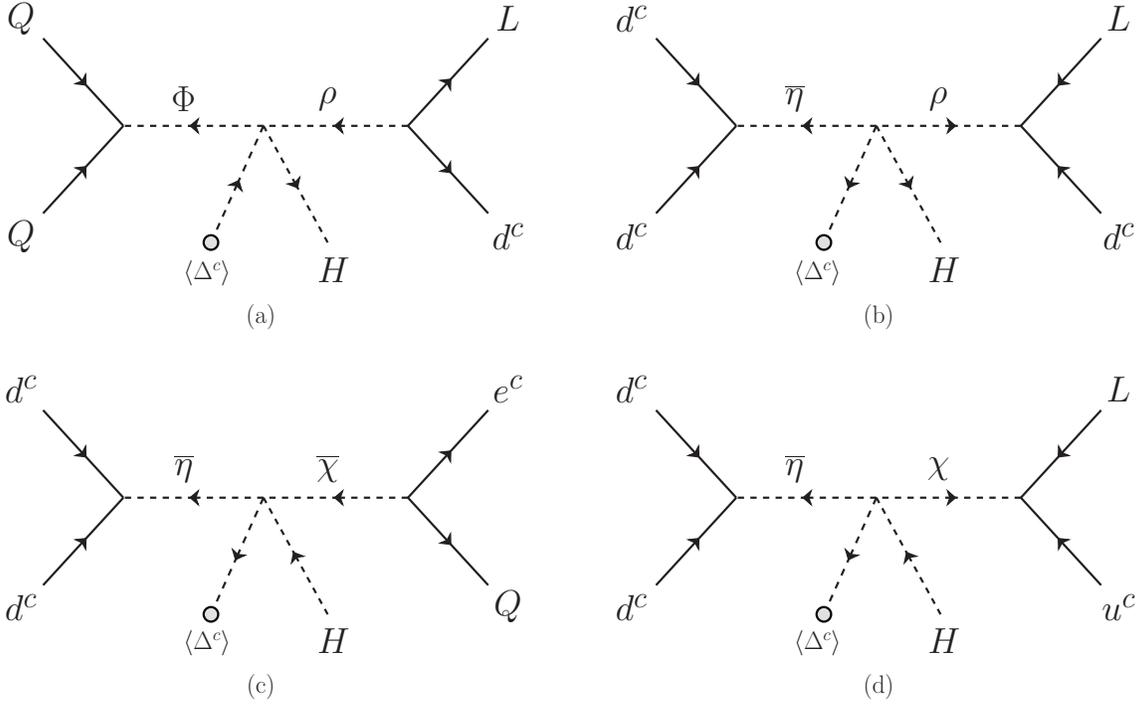}
	\caption{Effective $d=7$ baryon number violating operators obtained by
integrating our fields from the $120$. Here the SM quantum numbers of the scalar
fields are: $H(1,2,+1/2), \,\omega(3,1,-1/3),\, \rho(3,2,1/6),\, \eta(3,1,2/3),\,
\Phi(3,3,-1/3),$ and  $\chi(3,2,7/6).$}
\label{antisym}
\end{figure}

In Fig. \ref{antisym} we display the effective $d=7$ operators obtained by integrating
out the flavor antisymmetric $120$ coupling to fermions.  We see that four operators are induced this way.  As already noted, the required trilinear vertices to complete these diagrams arise from quartic couplings.  It is also possible to replace the $\rho$ fields in Fig. \ref{antisym} by
a $\rho$ field from the $\overline{126}$, in which case the sum of such diagrams would have no definite symmetry property in two of the flavor indices.

Note that all of the $d=7$ operators arising from Fig. \ref{sym} and \ref{antisym} respect $B-L$ symmetry, as can be seen by assigning $(B-L)(\Delta^c) = -2$.

It should be noted that operators ${\cal \tilde{O}}_7- {\cal \tilde{O}}_9$ of Eq. (\ref{dim7}) do not arise at tree level by
integrating out superheavy particles.  They can arise via loops, with suppressed strength.  In the estimate
of nucleon lifetime these operators play a subleading role, since the amplitude for the decay would be further
suppressed by the nucleon momentum (rather than the electroweak VEV for operators ${\cal \tilde{O}}_1-{\cal \tilde{O}}_6$).

What if the Higgs field employed for reducing the rank of $SO(10)$ is a $16_H$ rather than
a $126_H$?  In this
case the $d=7$ diagrams of Fig. \ref{sym} and \ref{antisym} would still arise, albeit in
a slightly different way.  The $(B-L)$ quantum numbers of the Higgs doublets $h(1,2,1/2)$ from
the $10_H$, $126_H$ and $120_H$ fields are all zero.  The $16_H$ contains a SM singlet filed
with $B-L = +1$ which acquires a GUT scale VEV.  It also contains  a $\overline{h}(1,2,-1/2)$ field with $B-L = -1$.
Similarly, $\overline{16}_H$ contains a SM singlet field with $(B-L) = -1$ and a $h(1,2,1/2)$ field with $(B-L) = +1$.
The trilinear scalar couplings of the type $16_H 16_H 10_H$ and $\overline{16}_H \overline{16}_H 10_H$
would mix the $B-L = 0$ Higgs doublet $h(1,2,1/2)$  from the $10_H$ and the
$h(1,2,1/2)$ Higgs from the
$\overline{16}_H$ which has $B-L = +1$.  The light SM Higgs doublet then would have no definite $B-L$ quantum
number.  The $(1,2,4)$ component of $\overline{16}_H$ under $G(2,2,4)$ contains the field $\rho^*(\overline{3},2,-1/6)$, and the $(2,1,\overline{4})$ of $\overline{16}_H$ under $G(2,2,4)$ contains $\omega(3, 1, -1/3)$, and thus the coupling $\rho^* \omega H$ is generated via the $\overline{16}_H
\overline{16}_H 10_H$ coupling.  One could also take $\omega(3,1,-1/3)$ from the $10_H$, $\rho^*(\overline{3},2,-1/6)$
from the $\overline{16}_H$ and $H(1,2,1/2)$ from the $\overline{16}_H$ to generate the $\rho^* \omega H$ coupling from
$\overline{16}_H \overline{16}_H 10_H$.
It should be noted that the $\rho^*(\overline{3}, 2, -1/6)$ field from the $\overline{16}_H$ does have Yukawa couplings to fermions, since in this
type of models the heavy Majorana masses of the $\nu^c$ fields arise from the couplings
$16_i 16_j \overline{16}_H \overline{16}_H$, and upon insertion of one VEV for the SM
singlet field here, the coupling of $\rho^*$ to fermions is realized.

We point out that the $\rho(3,2,1/6)$ field is partly in the Goldstone mode, associated with the breaking of $SO(10)$ to $SU(5)$.  However, since other fields such as $45$, $54$ or $210$
should be employed to complete the symmetry breaking down to the SM, one such physical
$\rho(3,2,1/6)$ will remain in the spectrum. This is because the $45$, $54$ and $210$
all contain a $\rho(3,2,1/6)$ field, and only one $\rho$ is absorbed by the gauge
multiplet.

\begin{figure}[h]
\centering
\includegraphics[scale=0.5]{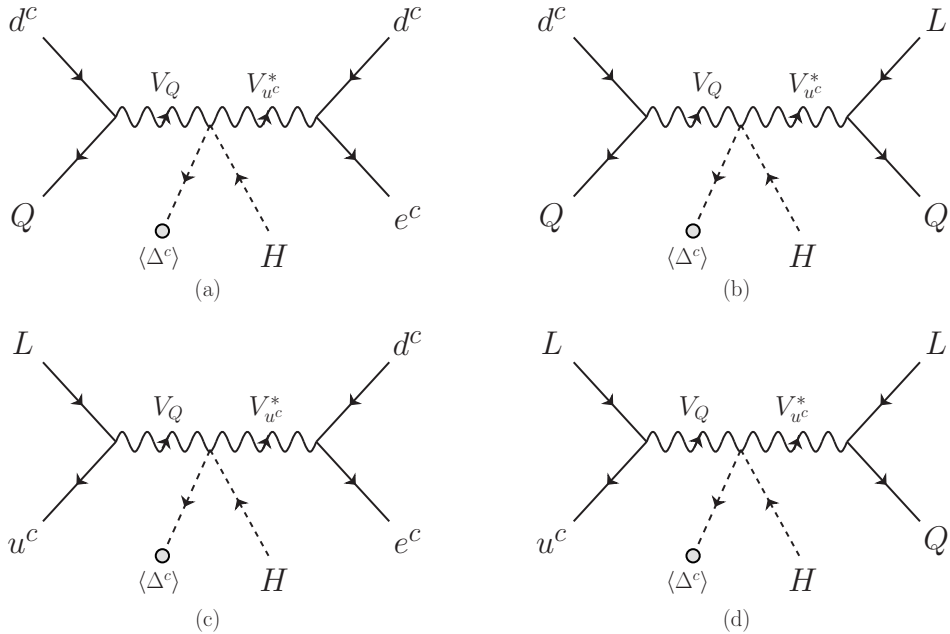}
\caption{Origin of $d=7$ baryon number violating operators via the exchange of
vector gauge bosons $V_Q(3,2,1/6)$ and $V_{u^c}(\overline{3}, 1, -2/3)$ of $SO(10)$.}
\label{gauge}
\end{figure}

The $d=7$ operators of Eq. (\ref{dim7}) can also arise by integrating out the gauge
bosons of $SO(10)$.  These vector operators are related by Fierz identities to the
scalar operators displayed in Eq. (\ref{dim7}).  The relevant diagrams are shown in
Fig. \ref{gauge}.  (Fig. \ref{gauge} (d) conserves $B$, but violates $L$ and $B-L$ \cite{leung}.
The effective operator from this diagram, after a Fierz rearrangement can be identified with
${\cal O}_7'$ of Eq. (\ref{lep}).)
In these diagrams $V_Q$ denotes the gauge boson with the SM quantum
numbers $(3,2,1/6)$ (same as the quark doublet $Q$), and $V_{u^c}$ is the gauge boson
that transforms as $(\overline{3},1,-2/3)$, the same as that of $u^c$ fermion.
In Fig. \ref{gauge}, each vertex conserves $B-L$ as it should, which can be seen
by assigning $(B-L) = (-2/3, -4/3)$ to $(V_Q, V_{u^c})$.  The covariant derivative
for the rank reducing field $126_H$ would contain the term $V_Q V_{u^c} H (\Delta^c)^\dagger$
which enters the $d=7$ operators.  When $16_H$ is used instead of the $126_H$, the
covariant derivative would contain a similar term, but now the $H$ and $\Delta^c$ fields
would carry $(B-L) = +1$ and $-1$ respectively.

\subsection{\boldmath{$B$}--violating \boldmath{$d=6$} superpotential operators in SUSY \boldmath{$SO(10)$}}

\begin{figure}[h]
\centering
\includegraphics[scale=0.5]{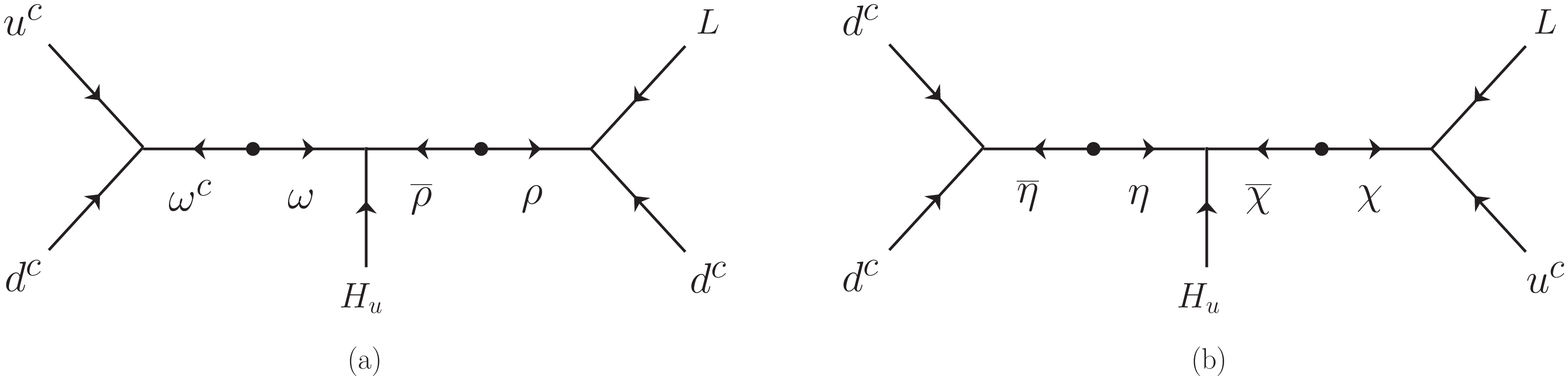}
\caption{Superfield diagrams for the generation of $d=6$ baryon number violating operator of Eq. (\ref{W}) in $SO(10)$.
Diagram (a) arises from integrating out colored fields from $10_H$ and $\overline{126}_H$, while diagram (b) arises from $120_H$.}
\label{susy}
\end{figure}

We have identified in Eq. (\ref{W}) a single holomorphic operator of dimension six in the superpotential which violates $B$ and $(B-L)$.
In Fig. \ref{susy} we show how this operator can arise in SUSY $SO(10)$.  Fig. \ref{susy} (a) is obtained by integrating
out colored fields in the $10_H$ and $\overline{126}_H$, while Fig. \ref{susy} (b) is obtained by integrating out such fields
from the $120_H$.  These are superfield diagrams, effectively generating $F$--terms in the Lagrangian.  Note that the effective
superpotential is antisymmetric in down-quark flavor in both diagrams (a) and (b).  This can be explicitly verified, by making
use of the $\epsilon\, u^c\, d^c\, \omega^c$ vertex in Fig. \ref{susy} (a) which is color antisymmetric.

In Fig. \ref{susy} (a), the vertex $L d^c \rho$ is contained in the Yukawa coupling of Eq. (\ref{Yuk126}) and the vertex
$u^c d^c \omega^c$ can be either from Eq. (\ref{Yuk10}) or from Eq. (\ref{Yuk126}).  The $\omega^c \omega$ and $\overline{\rho} \rho$
transitions occur through $(10_H)^2$ (or  $\overline{126}_H \, 126_H$) mass term, for example.  As for the vertex $\omega\, \overline{\rho}
\, H_u$ here is a concrete example.  The coupling $126_H \, 10_H\,  210_H$ is invariant, and contains such a term, where $\omega \subset 10_H$,
$\overline{\rho} \subset 126_H$ and $H_u \subset 210_H$.  Note that $H_u \subset 210_H$ carries $(B-L) = +2$, since it is part of the fragment
$(2,2,10)$ under $G(2,2,4)$.  Of course, the MSSM field $H_u$ in this case is an admixture of $h(1,2,1/2)$ from $10_H$,
$126_H$, $\overline{126}_H$ and $210_H$ and carries no definite $(B-L)$ charge.

As for Fig. \ref{susy} (b), the vertices $\chi\, L \, u^c$ and $\overline{\eta}\, d^c\, d^c$ are contained in the $120_H$ Yukawa coupling
of Eq. (\ref{Yuk120}).  The vertex $\eta \, \overline{\chi}\, H_u$ can arise from the superpotential coupling $(120_H)^2\, 210_H$, which
is invariant.

Suporpotential operators of Eq. (\ref{W}) can also arise in SUSY $SO(10)$ models which utilize low dimensional Higgs representations,
for e.g., $\{10_H + 10'_H+ 16_H + \overline{16}_H + 16'_H + \overline{16}'_H + 45_H\}$. Such models have been widely discussed in the literature \cite{raby,barr,bpw,bpt}.  While
$R$--parity is no longer automatic in these models (unlike in the case of $126_H$ models), it can be ensured by a discrete $Z_2$ symmetry which distinguishes $16_H$ from the chiral fermions $16_i$ and which remains unbroken.
The doublet--triplet splitting problem can be addressed without fine-tuning in these models
via the Dimopoulos--Wilczek mechanism \cite{dw}. Generation of fermion masses in these models relies on higher dimensional operators
such as $16_i 16_j \overline{16}_H \overline{16}_H$ which induce heavy Majorana masses for the $\nu^c$, and $16_i 16_j 16_H 16'_H$
which induce lighter family masses and CKM mixings.  ($16_H$ and $\overline{16}_H$ acquire GUT scale VEVs along their SM singlet
components, while $16'_H$ and $\overline{16}'_H$ do not.  The $SU(2)_L$ doublet components of $16_H$ and $16'_H$ acquire weak scale VEVs,
see for e.g., discussions in Ref. \cite{bpt}.)
To see the origin of Eq. (\ref{W}) in such models, consider generation
of Fig. \ref{susy} (a).  The vertex $L \, d^c \, \rho$ arises from the Yukawa coupling $16_i 16_j 16_H 16'_H$ after a GUT scale VEV
is inserted for the $16_H$ (note that $16'_H$ contains a $\rho$ field),
and the vertex $u^c \, d^c \, \omega^c$ is contained in the coupling $16_i 16_j 10_H$.  Now, $\omega^c$
can convert itself into $\omega \subset 10'_H$ via the coupling $10_H \,10'_H\, 45_H$ when the $(B-L)$--preserving VEV of $45_H$
is inserted, while the $\rho$ can transition into $\overline{\rho} \subset
45_H$ via the coupling  $16'_H \, \overline{16}_H \, 45_H$ with the insertion of a $\overline{16}_H$ VEV.  The $\overline{\rho} \, \omega \, H_u$ term is contained in the coupling $10_H \, 10'_H\, 45_H$, which completes the diagram. Explicit models \cite{bpt} contain all these terms necessary
for generating the $d=6$ superpotential operator.

\section{\boldmath{$(B-L)$}--violating nucleon decay rates in \boldmath{$SO(10)$}}

Before discussing the rates for $(B-L)$--violating nucleon decay, let us note that certain scalar bosons and
certain gauge bosons would induce the more dominant $d=6$ baryon number violating operators of Eq. (\ref{dim6}).
Scalar bosons with SM quantum numbers $\omega(3,1,-1/3)$, $\Phi(3,3,-1/3)$ and $\delta(3,1,-4/3)$ and
vector gauge bosons with SM quantum numbers $X(3,2,-5/6)$ and $V_Q(3,2,1/6)$ can can induce these $(B-L)$ preserving
$d=6$ operators.  Current nucleon lifetime limits restrict the masses of these gauge bosons to be larger than about
$10^{15}$ GeV, and those for the scalar bosons to be heavier than about $10^{11}$ GeV (for Yukawa couplings of
order $10^{-3}$).  Since some of the $d=7$ operators arise in $SO(10)$ via the exchange of these particles, the
above limits have to be met in our estimate of $d=7$ decay rates.

We proceed to estimate the lifetime of the nucleon for its $(B-L)$--violating decays
in $SO(10)$ a straightforward way.\footnote{$(B+L)$--preserving nucleon decay has been studied
in the context of $R$--parity breaking SUSY in Ref. \cite{vissani}.}  The diagrams of Fig. \ref{sym} lead to the estimate
\begin{equation}
\Gamma (n \rightarrow e^-\pi^+)^{\rm Fig. \ref{sym}} \approx \frac{|Y^*_{QQ\omega} Y_{L d^c \rho}|^2}{64 \pi}(1+D+F)^2 \frac{\beta_H^2 m_p}{f_\pi^2}\left(\frac{\lambda v v_R}{M^2_\rho} \right)^2
\frac{1}{M_\omega^4}~.
\label{tau}
\end{equation}
Here we have defined the Yukawa couplings of $\omega$ and $\rho$ fields appearing in Fig. \ref{sym} to be
$Y^*_{QQ\omega}$ and  $Y_{L d^c \rho}$.  These couplings are linear combinations of the Yukawa coupling matrices $h$, $f$ and $g$ of
Eqs. (\ref{Yuk10})-(\ref{Yuk120}) with flavor indices corresponding to the first family fermions.
The factors $D$ and $F$ are chiral Lagrangian factors, $D \simeq 0.8$
and $F \simeq 0.47$.  $\beta_H \simeq 0.012~{\rm GeV}^3$ is the nucleon decay matrix element \cite{aoki},
$v_R \equiv \left\langle \Delta^c\right\rangle$, and $v\equiv \left \langle H^0 \right \rangle = 174$ GeV.
We have defined the trilinear coupling of Fig. \ref{sym} to have a coefficient $\lambda v_R$. As expected,
the rate is suppressed by six powers of inverse mass, owing to the higher dimensionality of the effective operator.
The mass of $\omega(3,1,-1/3)$ is constrained to be relatively large, as it
mediates $d=6$ nucleon decay.  For $Y\approx 10^{-3}$, $M_\omega > 10^{11}$ GeV
must be met from the $d=6$ decays.  As an illustration, choose $Y_{QQ\omega}=Y_{L d^c \rho} =  10^{-3}$,
$M_\omega = 10^{11}$ GeV, $M_\rho = 10^8$ GeV, and $\lambda v_R = 10^{11}$ GeV in Eq. (\ref{tau}).
This choice would result in $\tau_n \approx  3 \times 10^{33}$ yrs.  Such a spectrum is motivated by
the intermediate symmetry $G(2,2,4) = SU(2)_L \times SU(2)_R \times SU(4)_C$ (without discrete Parity), which is found  to be realized at $M_I \approx 10^{11}$ GeV from gauge coupling unification \cite{chang}.
As a second example, take $M_\rho = 10^6$ GeV, $M_\omega = 10^{16}$ GeV, $\lambda v_R = 10^{16}$ GeV, $Y_{QQ\omega}=Y_{L d^c \rho} = 3 \times 10^{-3}$. This choice of spectrum leads to $\tau_n \approx 4 \times  10^{33}$ yrs.

\begin{figure}[h]
\centering
	\includegraphics[scale=0.73]{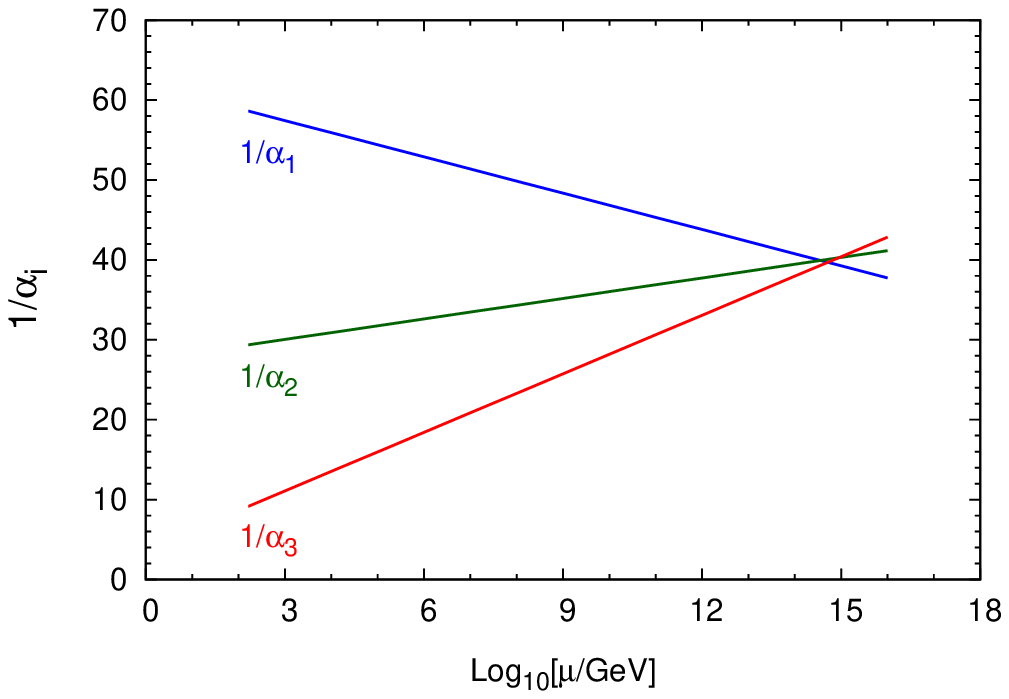}
\includegraphics[scale=0.73]{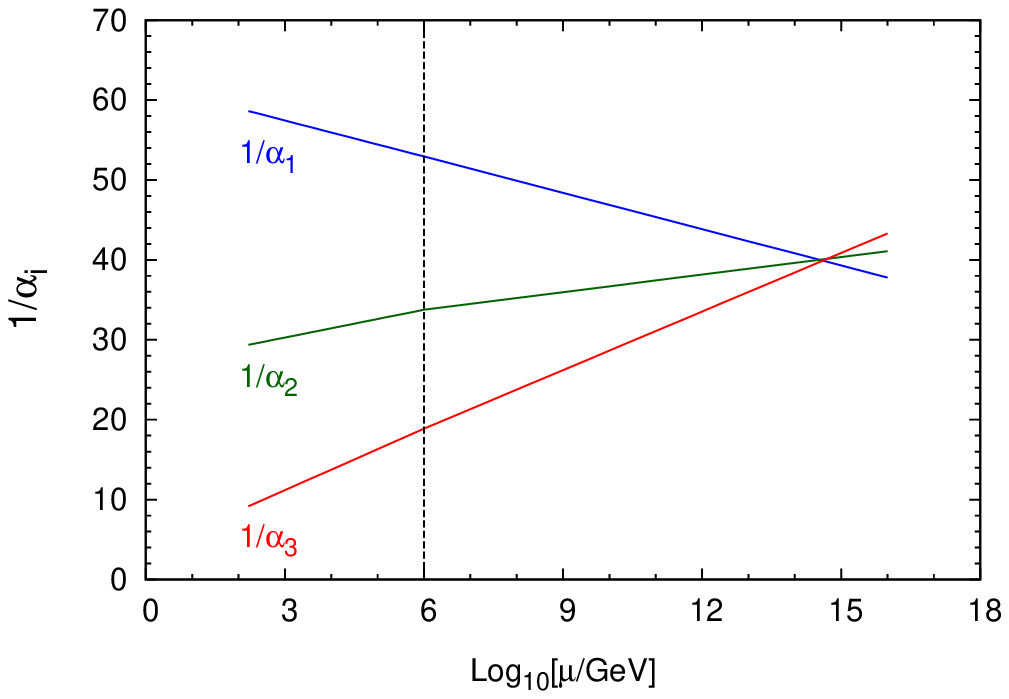}
	\caption{Unification of the three SM gauge couplings obtained with a light
$\rho(3,2,1/6)$ and two $(1,3,0)$ scalar multiplets at $M_I= 10^6$ GeV (right panel).
   The left panel corresponds to having one $\rho(3,2,1/6)$ and one $(1,3,0)$ scalar fields
   at the weak scale.}
	\label{unif}
\end{figure}

This second choice for the spectrum can be motivated as follows.  The unification of gauge couplings
may occur without any particular intermediate symmetry, but with certain particles surviving to an
intermediate scale. We have found two examples of this type where the $d=7$ nucleon decay is within
observable range.  Suppose the $\rho(3,2,1/6)$ particle, along with a pair of $(1,3,0)$ scalar particles
(contained in the $45_H$, $54_H$ or $210_H$ needed for symmetry breaking) survive down to $M_I = 10^6$ GeV.  The three SM gauge couplings
are found to unify at a scale $M_X \approx 10^{15}$ GeV in this case.  This is shown in Fig. \ref{unif}
right panel.  This figure is obtained by the one-loop renormalization group evolution of the SM gauge couplings above $M_I$
with beta function coefficients $b_i = (62/15, -2, -20/3)$, where $dg_i/d{\rm ln}\mu = b_ig_i^3/(16 \pi^2)$.
Since the $d=6$ nucleon decay lifetime would also be near $\tau(p \rightarrow e^+ \pi^0)\approx
10^{34}$ yrs with such a unification scale, this scenario would predict observable rates for both the
$(B-L)$--conserving and $(B-L)$--violating nucleon decay modes.  Another possible scenario for consistent gauge coupling
unification is to assume that $\rho(3,2,1/6)$ and one $(1,3,0)$ scalar multiplet survive down to the weak scale.
Again, the gauge couplings unify around $M_X \approx 10^{15}$ GeV, as shown in the left panel of Fig. \ref{unif}.
Here the $\beta$--function coefficients used are $b_i = (62/15,-7/3, -20/3)$, appropriate for this spectrum.
The estimate $\tau(n \rightarrow e^- K^+) \approx 10^{33}$ yrs would follow, as in the second example above, but
now with $Y_{QQ\omega}=Y_{L d^c \rho} = 10^{-4}$.

Nucleon decay rates arising from diagrams of Fig. \ref{antisym} are similar, but with a significant
difference.  Take for example Fig. \ref{antisym} (b) with the $\eta \rho^* H$ trilinear vertex.
Neither the $\eta(3,1,2/3)$ nor the $\rho(3,2,1/6)$ field would mediate
$d=6$ nucleon decay, and may be considerably lighter than the GUT scale.  In non--supersymmetric
$SO(10)$ models it is natural that some scalars survive down to an intermediate scale.
Typically the intermediate scale is of order $10^{10}- 10^{12}$ GeV if the associated
symmetry contains $SU(4)_C$.  Thus it is possible that both $\eta$ and $\rho$ have
masses of order $10^{10}$ GeV.  Nucleon decay rate is now estimated to be
\begin{equation}
\Gamma (n \rightarrow e^-\pi^+)^{\rm Fig. \ref{antisym}} \approx \frac{|Y_{L d^c \rho} Y_{d^c d^c\overline{\eta}}|^2}{64 \pi}(1+D+F)^2 \frac{\beta_H^2 m_p}{f_\pi^2}\left(\frac{\lambda v v_R}{M^2_\rho} \right)^2
\frac{1}{M_\eta^4}~.
\end{equation}
Here $\lambda v_R$ is defined as the coefficient of the trilinear scalar vertex.
For $M_\rho = M_\eta = 10^{10}$ GeV, $v_R = 10^{11}$ GeV, $Y = 10^{-2}$,
$\tau_n \approx 3 \times 10^{33}$ yrs, which is in the observable range.

The gauge boson exchange diagrams of Fig. \ref{gauge} would also induce $(B-L$)--violating nucleon decay.
However, we find the rates for these decays to be suppressed.  Take Fig. \ref{gauge} (a) for example.
The vector gauge boson $V_Q$ must have mass of order $10^{15}$ GeV, since it mediates $d=6$ nucleon decay,
while $V_{u^c}$ could be much lighter as it does not have $B$--violating interaction by itself.  The
$V_Q V_{u^c} H$ vertex necessary for connecting the $d=7$ diagram has a coefficient of order $g^2 v_R$,
while the mass of $V_{u^c}$ is of order $g v_R$.  The amplitude for $d=7$ nucleon decay arising
from Fig. \ref{gauge} is then given by $A \approx (g^4 v_R v)/(M_{V_Q}^2 M_{V_{u^c}}^2) \approx (g^2/M_{V_Q}^2)(v/v_R)$.
This amplitude is a factor $(v/v_R)$ smaller compared to the standard $d=6$ nucleon decay amplitude originating
from GUT scale gauge bosons. Unless the $(B-L)$--breaking scale $v_R$ is close to the weak scale $v$, not a
likely scenario based on gauge coupling unification, nucleon
lifetime for $(B-L)$--violating modes from these diagrams would be beyond the reach of experiments.  However, as we shall see in the next
section, the $(B-L)$--violating decays of the gauge boson can naturally explain the observed baryon asymmetry
of the universe.

As for the supersymmetric diagram of Fig. \ref{susy}, the $d=6$ operator in the superpotential Eq. (\ref{W}) has a strength
$A = (Y_{u^c d^c \omega^c}\, Y_{L d^c \rho}\, Y_{\omega \overline{\rho} H_u})/(M_\omega M_\rho)$.  The Lagrangian would
contain a term $\tilde{d^c} \tilde{s^c} (u^c \nu) v_u$, upon insertion of the VEV of $H_u$.  The scalars can be converted
to $d$ and $s$ quarks via a gluino loop, and would result in the following estimate for processes such as $n \rightarrow \nu K^0$:
\begin{equation}
\Gamma (n \rightarrow \nu K^0)^{\rm Fig. \ref{susy}} \approx \frac{\left|Y_{u^c d^c \omega^c}\, Y_{L d^c \rho}\, Y_{\omega \overline{\rho}H_u}\right|^2}{64\pi}
(1+D+F)^2 \frac{\beta_H^2 m_p}{f_\pi^2}\left(\frac{v_u}{M_S} \right)^2\left(\frac{\alpha_s}{\pi} \right)^2
\frac{1}{M^2_\omega M^2_\rho}~.
\end{equation}
Here $M_S$ is a typical SUSY breaking mass scale.  For $Y_{u^c d^c \omega^c}= Y_{L d^c \rho} = 10^{-2}$, $Y_{\omega \overline{\rho}H_u}= 1.0$,
$M_\rho =10^{9}$ GeV, $M_\omega=10^{16}$ GeV, $M_S = 1$ TeV, one obtains $\tau(n \rightarrow \nu K^0) \approx 10^{34}$ yrs.

As for the consistency of such an intermediate scale mass for $\rho$ with gauge coupling unification in SUSY models,
we note that $\rho(3,2,1/6)$ forms a complete $SU(5)$
$10$--plet along with a $(\overline{3},1,-2/3)$ and a $(1,1,1)$ fields.  If these fields also have an intermediate scale mass,
unification of gauge coupling would work as in the MSSM.

It should be noted that in SUSY models, the decay $n \rightarrow e^- K^+$ would be suppressed, since the VEV of $H_u$ picks
a neutrino field in Eq. (\ref{W}).  Discovery of $n \rightarrow e^- K^+$ decay would thus hint at a deeper non--supersymmetric
dynamics.

\section{Baryogenesis at the GUT epoch}

We now proceed to the computation of the baryon asymmetry of the universe induced at the GUT epoch. The $(B-L)$--violating decays
of the scalars $\omega(3,1,-1/3)$ and $\eta(3,1,2/3)$ and of the vector gauge boson $V_Q(3,2,1/6)$ will be used to illustrate
the mechanism.  We shall see that in each case, the out of equilibrium condition can be satisfied, and that there is enough
CP violation.  These decays generate an asymmetry in $(B-L)$, which is not destroyed by the effective interactions induced
by the electroweak sphalerons, and would survive to low temperatures.  This is in contrast with the induced baryon asymmetry
in the $(B-L)$--preserving decays of GUT scale scalars and gauge bosons in unified models such as $SU(5)$, which is washed out
by the spharleron interactions.

\subsection{\boldmath{$(B-L)$} asymmetry in \boldmath {$\omega \rightarrow \rho H^*$} decay}

\begin{figure}[h]
\centering
	\includegraphics[scale=0.5]{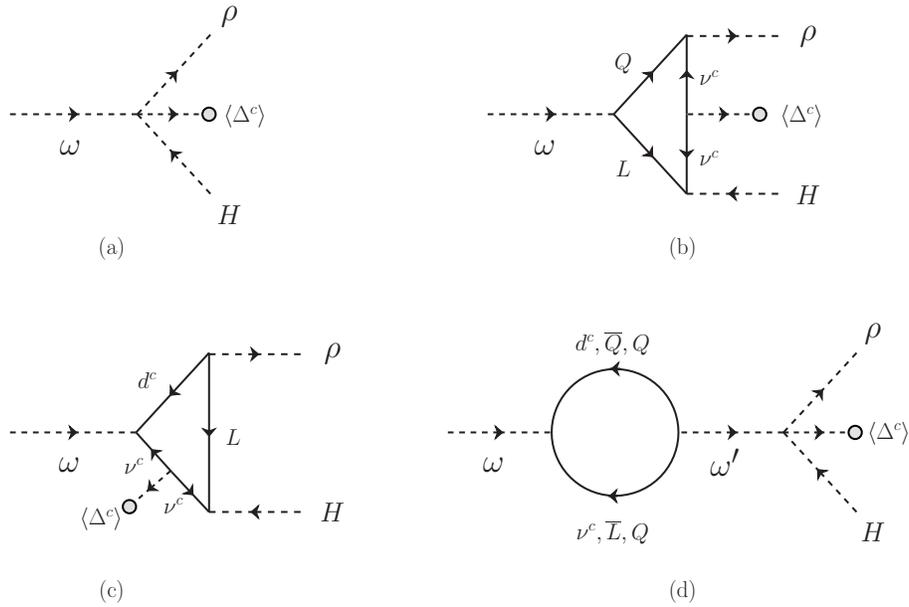}
	\caption{Tree--level diagram and one--loop corrections responsible for generating $(B-L)$ asymmetry in $\omega$ decay.}
	\label{baryo1}
\end{figure}

We begin with the $(B-L)$--violating decay of the scalar $\omega(3,1,-1/3)$ which is assumed to have a mass of order the GUT scale.
($B-L$ asymmetry in decays of specific heavy particles has recently been discussed in Ref. \cite{maekawa}.)
To be concrete, we shall work in the framework of non--supersymmetric $SO(10)$, although our results would hold for SUSY $SO(10)$
as well, with some minor modifications.  We identify $\omega$ to be the lightest of the various $\omega_i(3,1,-1/3)$ scalar fields in
the $SO(10)$ theory.  The Yukawa couplings of Eq. (\ref{Yuk10})-(\ref{Yuk120}) imply that $\omega$ (which is
in general a linear combination of
$\omega_i, \omega^{c*}_i$ from the $10_H$ and $\overline{126}_H$ and $120_H$ fields) has two--body decays into fermions of the type
$\omega \rightarrow \overline{Q}\, \overline{Q},\,\overline{u^c}\,\overline{e^c},\,\overline{\nu^c}\, \overline{d^c},\,u^c\, d^c,\, Q\,L$.
These decays preserve $(B-L)$, as can be seen by assigning $(B-L)(\omega) = -2/3$.  Now, $\omega$ also has a two--body scalar decay,
$\omega \rightarrow \rho H^*$ as shown in Fig. \ref{baryo1} (a), which uses the $(B-L)$ breaking VEV of $\Delta^c$.
The scalar field $\rho$ has two--body fermionic decays of the type
$\rho \rightarrow \overline{L}\, \overline{d^c},\,\nu^c \,Q$ (the latter if kinematically allowed), which define $(B-L)$ charge of
$\rho$ to be $+4/3$.  Thus the decay $\omega \rightarrow \rho H^*$ would violate $(B-L)$ by $-2$ (recall that $H$ has zero $(B-L)$ charge).

Focussing on the $(B-L)$--violating decay $\omega \rightarrow \rho H^*$, we define a $(B-L)$ asymmetry parameter $\epsilon_{B-L}$ as follows.
Let the branching ratio for $\omega \rightarrow \rho H^*$ be $r$ which produces a net $(B-L)$ number of $4/3$, and that for $\omega^* \rightarrow \rho^* H$ be $\overline{r}$, with net $(B-L) = -4/3$.  The branching ratio for the two--fermion decays
$\omega \rightarrow ff$ is then $(1-r)$ which has  $(B-L) = -2/3$, and that
for $\omega^* \rightarrow \overline{f}\, \overline{f}$ is $(1-\overline{r})$ which has $(B-L) = 2/3$.
Thus in the decay of a $\omega + \omega^*$ pair, a net $(B-L)$ number, defined as $\epsilon_{B-L}$, is induced, with
\begin{eqnarray}
\epsilon_{B-L}  &\equiv & (B-L)_\omega + (B-L)_{\omega^*} = \frac{4}{3}(r-\overline{r}) -\frac{2}{3}\left\{(1-r) - (1-\overline{r})  \right\}
\nonumber \\
&=& 2(r-\overline{r})~.
\end{eqnarray}
Note that an interplay between the $(B-L)$--conserving decays and the $(B-L)$--preserving decays of $\omega$ is necessary for inducing
this asymmetry.  In addition, CP violation is required, otherwise $r=\overline{r}$ and thus $\epsilon_{B-L} = 0$.  Nonzero
$\epsilon_{B-L}$ also requires a loop diagram which has an absorptive part. All these conditions are realized in $SO(10)$ models.
The loop diagrams for $\omega\rightarrow \rho H^*$ are shown in Fig. \ref{baryo1} (b)-(d), which involve the exchange of fermions.
Since $\omega$ can also decay to two on--shell fermions, these loop diagrams have absorptive parts.  Note that Fig. \ref{baryo1} (b)-(c)
are proportional to the Majorana masses for the $\nu^c$ fields, while Fig \ref{baryo1} (d) is not.  These diagrams have CP violating
phases, ensuring a nonzero $\epsilon_{B-L}$.

We evaluate Fig. \ref{baryo1} in a basis where the Majorana mass matrix of the $\nu^c$ fields is diagonal and real.  The contribution of
Fig. \ref{baryo1} (b) to $\epsilon_{B-L}$ is found to be
\begin{equation}
\epsilon_{B-L}^{(b)} = -\frac{1}{\pi} {\rm Im} \left[\frac{{\rm Tr}\{Y^\dagger_{QL \omega^*} \, Y_{Q \nu^c \overline{\rho}}\,M_{\nu^c} \, F_1(M_{\nu^c})\, Y_{\nu^c L H} )\} \, \lambda v_R}{|\lambda v_R|^2} \right] {\rm Br}.
\label{asymb}
\end{equation}
Here we have defined the trilinear scalar vertex of Fig. \ref{baryo1} (a) to have a coefficient $\lambda v_R$ in the Lagrangian.
$Y_{Q \nu^c \overline{\rho}}$ is
the Yukawa coupling matrix corresponding to the coupling $Q \,\nu^c\, \overline{\rho}$, etc.
$M_{\nu^c}$ is the diagonal and real mass matrix of $\nu^c$ fields.
${\rm Br}$ stands for the branching ratio
${\rm Br}(\omega \rightarrow \rho H^*)$.  A factor of $2$ has been included here for the two $SU(2)_L$ final states in the decay.
The function $F_1(M_j)$ is defined as
\begin{equation}
F_1(M_j) = {\rm ln}\left(1+ \frac{M_\omega^2}{M_j^2}\right) + \Theta\left(1- \frac{M_j^2}{M_\rho^2}\right)\,\left(1- \frac{M_j^2}{M_\rho^2}\right)
\end{equation}
with $M_j$ denoting the mass of $\nu^c_j$.  Here $\Theta$ stands for the step function, signalling additional ways of cutting the
diagram when $M_j < M_\rho$ in Fig. \ref{baryo1} (b).

Fig. \ref{baryo1} (c) yields the following contribution to $\epsilon_{B-L}$:
\begin{equation}
\epsilon_{B-L}^{(c)} =  \frac{1}{\pi} {\rm Im} \left[\frac{{\rm Tr}\{Y_{d^c \nu^c \omega} \,Y^\dagger_{d^c L \rho} \,\, Y_{\nu^c L H} \, M_{\nu^c}\, F_2(M_{\nu^c})\}\, \lambda v_R}{|\lambda v_R|^2} \right]{\rm Br},
\label{asymc}
\end{equation}
where $F_2(M_j)$ is defined as
\begin{equation}
F_2(M_j) = {\rm ln}\left(1+ \frac{M_\rho^2}{M_j^2}\right) + \Theta\left(1- \frac{M_j^2}{M_\omega^2}\right)\,\left(1- \frac{M_j^2}{M_\omega^2}\right).
\end{equation}

Fig. \ref{baryo1} (d) arises because in any realistic $SO(10)$ model there are at least two $\omega$ fields.  The heavier $\omega$
field is denoted as $\omega'$.  In principle one can sum over all such $\omega'$ contributions, but here we have kept only one such
$\omega'$ field.  Its contribution to $\epsilon_{B-L}$ is found to be
\begin{equation}
\epsilon_{B-L}^{(d)} = \frac{1}{\pi} {\rm Im} \left[\frac{{\rm Tr}\{Y^\dagger_{d^c \nu^c \omega'} \, Y_{d^c \nu^c \omega}
\, F_3(M_{\nu^c}) \}  (\lambda'v_R)^*(\lambda v_R)}{|\lambda v_R|^2} \right]  {\rm Br},
\label{asymd}
\end{equation}
where $F_3(M_j)$ is defined as
\begin{equation}
F_3(M_j) =  \left(\frac{M_\omega^2 - M_j^2}{M_\omega^2-M_{\omega'}^2}\right) \Theta\left(1- \frac{M_j^2}{M_\omega^2}\right)\,\left(1- \frac{M_j^2}{M_\omega^2}\right)~.
\end{equation}
This contribution, which is non-vanishing even in the limit of vanishing $M_j$, requires $M_j < M_\omega$.
Here we have defined the trilinear coupling $\rho^* \omega' H$ to have a coefficient $\lambda' v_R$ in the Lagrangian.  We have also
assumed that $M_\omega - M_\omega' \gg \Gamma_\omega$, so that there is no resonant enhancement for the decay.  (When $\omega$ and
$\omega'$ are nearly degenerate in mass, such a resonant enhancement is possible.  In this case, the expression for the decay rate
will be smoothened by the width of these particles.  Appropriate expressions in this case can be found in
Ref. \cite{pilaftsis}.)

The branching ratio factor ${\rm Br} = {\rm Br}(\omega \rightarrow \rho H^*)$  appearing in Eqs. (\ref{asymb}), (\ref{asymc}), (\ref{asymd})
can be estimated as follows.  For this purpose let us assume that $\omega$ is the field $\omega$ from $10_H$ with Yukawa
couplings as given in Eq. (\ref{Yuk10}).  The partial widths for the decays $\Gamma_1(\omega \rightarrow \rho H^*)$ and $\Gamma_2(\rho
\rightarrow ff)$ are then given by
\begin{equation}
\Gamma_1(\omega \rightarrow \rho H^*) = \frac{|\lambda v_R|^2}{8 \pi M_\omega}\left(1-\frac{M_\rho^2}{M_\omega^2}  \right),~~~~~
\Gamma_2(\omega \rightarrow ff) = \frac{{\rm Tr}(h^\dagger h) }{4 \pi} M_\omega~.
\label{partial}
\end{equation}
In the expression for $\Gamma_2$ we have assumed that $\nu^c$ is much lighter than $\omega$.  In terms of these partial widths,
the branching ratio that appears in Eqs. (\ref{asymb}), (\ref{asymb}), (\ref{asymd}) is given as ${\rm Br} = \Gamma_1/(\Gamma_1 + \Gamma_2)$.
To get a feeling for numbers, let us choose a realistic set of parameters:
$M_\omega = 10^{16}$ GeV, $h_{33} = 0.6$ (corresponding to the top quark Yukawa
coupling at GUT scale) with other $h_{ij}$ negligible, and $\lambda v_R = (10^{14},\,10^{15},\,10^{16})$ GeV.  This would
correspond to ${\rm Br} = (1.4 \times 10^{-4},\, 1.4 \times 10^{-2},\,0.58)$, which shows a strong dependence on $\lambda v_R$.

The total $(B-L)$ asymmetry in $\omega \rightarrow \rho H^*$ and its conjugate decay is given by
\begin{equation}
\epsilon_{B-L} = \epsilon_{B-L}^{(b)} +  \epsilon_{B-L}^{(c)} +  \epsilon_{B-L}^{(d)}~.
\end{equation}
This will result in the baryon to entropy ratio $Y_B$ given by
\begin{equation}
Y_B \equiv \frac{n_B-n_{\overline{B}}}{s} = \frac{\epsilon_{B-L}}{g_*} d~,
\label{eta}
\end{equation}
where $g_*$ is the total number of relativistic degrees of freedom at the epoch when these decays occur.
In our present example $g_* = 130$ which includes the SM particles and the $\rho$ and $\omega$ scalar fields.
The factor $d$ in Eq. (\ref{eta}) is the dilution factor which takes into account back reactions that would
partially wash out the induced baryon asymmetry.  $d$ is determined by solving the Boltzmann equations numerically, but
simple analytic approximations are available suitable to the present setup.  Defining a ratio
\begin{equation}
K = \left. \frac{\Gamma(\omega \rightarrow \rho H^*)} {2{\rm H}}\right|_{T=M_\omega},
\end{equation}
where ${\rm H}$ is the Hubble expansion rate,
\begin{equation}
{\rm H} = 1.66\, g_*^{1/2} \frac{T^2}{M_{\rm Pl}},
\end{equation}
the dilution factor can be written as \cite{kolb}
\begin{equation}
d\simeq \left\{\begin{array}{ll}
 1 & ~~~(K < 1) \\
 \frac{0.3}{K ({\rm ln}\,K)^{0.6}} & ~~~(K \gg 1).
\end{array} \right.
\end{equation}
These approximations work well for  $K < 100$ or so, beyond which $d$ would be exponentially suppressed.
For $M_\omega = 10^{16}$ GeV, $\lambda v_R = (10^{14},\,10^{15},\,10^{16})$ GeV, we find $K = (1.3 \times 10^{-4},\,1.3 \times
10^{-2},\, 1.23)$, with the corresponding dilution factors being $d= (1.0,\, 1.0,\,0.63)$.  Thus we see that there is
not much dilution with this choice of parameters, although ${\rm Br} = (1.4 \times 10^{-4},\, 1.4 \times 10^{-2},\,0.58)$
can become small for smaller values of $\lambda v_R$.  If we choose $M_\omega = 10^{15}$ GeV instead, and vary
 $\lambda v_R = (10^{14},\,10^{15},\,10^{16})$ GeV, then we find $K = (0.12,\,12.3,\,1230)$ and the corresponding
 dilution factors to be $d= (1.0,\,1.4 \times 10^{-2},\, 7.5 \times 10^{-5})$, with ${\rm Br}= (1.3 \times 10^{-2},\,0.58,\,1.0)$.

Although the electroweak sphaleron interactions would not wash away the GUT scale induced $(B-L)$ asymmetry, partial
wash--out can occur via the $(B-L)$--violating interactions of the right--handed neutrinos.  This is possible because the
$\nu^c$ fields acquire $(B-L)$--violating Majorana masses, and their interactions with the Higgs field and the lepton fields can
erase part of the GUT--induced asymmetry.  In some cases it may be desirable to have partial wash--out.
One can also prevent any wash--out by decoupling the $\nu^c$ fields at the same temperature as the $\omega$ field
(which would happen if $M_{\nu^c} \sim M_\omega$).  For the surviving light $\nu^c$ fields, the condition
$Y^2/(8\pi) \ll 1.66 g_*^{1/2}(M_{\nu^c}/M_{\rm Pl})$ would guarantee no further wash--out, where $Y$ is
the Dirac Yukawa coupling of the light $\nu^c$ field.

While we do not present the calculation of $Y_B$ in the supersymmetric version of the $SO(10)$ model, results
in that case would be similar to the non--SUSY case.  More diagrams contribute to the generation of $(B-L)$ asymmetry
with superparticle decays included.  The number of relativistic degrees of freedom $g_*$ would also double in this
case.

\subsection{\boldmath{$(B-L)$} asymmetry in \boldmath {$\eta \rightarrow \rho H$} decay}

\begin{figure}[h]
\centering
	\includegraphics[scale=0.5]{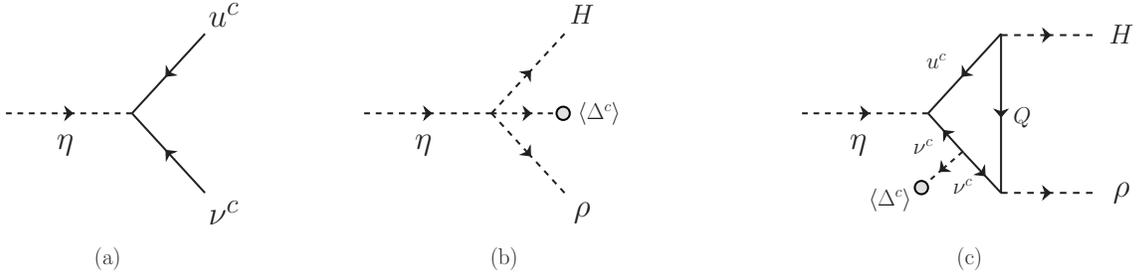}
	\caption{Tree--level and one--loop level diagrams responsible for inducing $(B-L)$ asymmetry in $\eta \rightarrow \rho H^*$ decay.}
	\label{baryo2}
\end{figure}

The $(B-L)$ asymmetry induced in the decay $\eta(3,1,2/3) \rightarrow \rho H$ is similar to the one induced in
the decay $\omega \rightarrow \rho H^*$.  The tree--level $(B-L)$--violating decay and the one--loop corrections
are shown in Fig. \ref{baryo2}.  $\eta$ has a fermionic decay mode $\eta \rightarrow \overline{u^c}\,\overline{\nu^c}$ shown
in Fig. \ref{baryo2} (a),
which can be used to define its $(B-L)$ quantum number as $(B-L)(\eta) = -2/3$.  The decay $\eta \rightarrow H \rho$ (Fig.
\ref{baryo2} (b)) would then violate $(B-L)$ by $+2$ units, since $(B-L)(\rho) = 4/3$ obtained from the decays $\rho \rightarrow
\overline{L}\,\overline{d^c}, \nu^c Q$.  The one--loop correction to this decay is shown in Fig. \ref{baryo2} (c), which is
evaluated in analogy to Fig. \ref{baryo1} (c) to be
\begin{equation}
\epsilon_{B-L}^{(\eta)} =  \frac{1}{\pi} {\rm Im} \left[\frac{{\rm Tr}\{Y_{u^c \nu^c \eta} \,Y^\dagger_{u^c Q H} \,\, Y_{\nu^c Q \rho} \, M_{\nu^c}\, \hat{F}_2(M_{\nu^c})\}\, \lambda v_R}{|\lambda v_R|^2} \right]{\rm Br}.
\label{asymeta}
\end{equation}
Here the function $\hat{F}_2(M_j)$ is defined as
\begin{equation}
\hat{F}_2(M_j) = {\rm ln}\left(1+ \frac{M_\rho^2}{M_j^2}\right) + \Theta\left(1- \frac{M_j^2}{M_\eta^2}\right)\,\left(1- \frac{M_j^2}{M_\eta^2}\right),
\end{equation}
and  $\lambda v_R$ identified as the coefficient of the trilinear vertex $\rho^* \eta H^*$.

A noteworthy feature of baryon asymmetry generated in the decay $\eta \rightarrow \rho H^*$ is that both the scalars $\eta$ and $\rho$ can
be relatively light, since they do not induce $d=6$ baryon number violating operators of Eq. (\ref{dim6}).
In supersymmetric extensions of $SO(10)$, this allows for the possibility that the gravitino abundance
problem of supergravity models can be evaded.  Typically in supergravity models, the reheat temperature after inflation
is required to be $T_{\rm reheat} < 10^8$ GeV, in order to sufficiently dilute the gravitino abundance
in the universe. If $M_\eta,\,M_\rho < 10^8$ GeV, this requirement would be compatible with the $(B-L)$ asymmetry
generation.

\subsection{\boldmath{$(B-L)$} asymmetry in vector gauge boson decay \boldmath{$V_Q \rightarrow V_{u^c}^* \,H^*$}}

\begin{figure}[h]
\centering
	\includegraphics[scale=0.5]{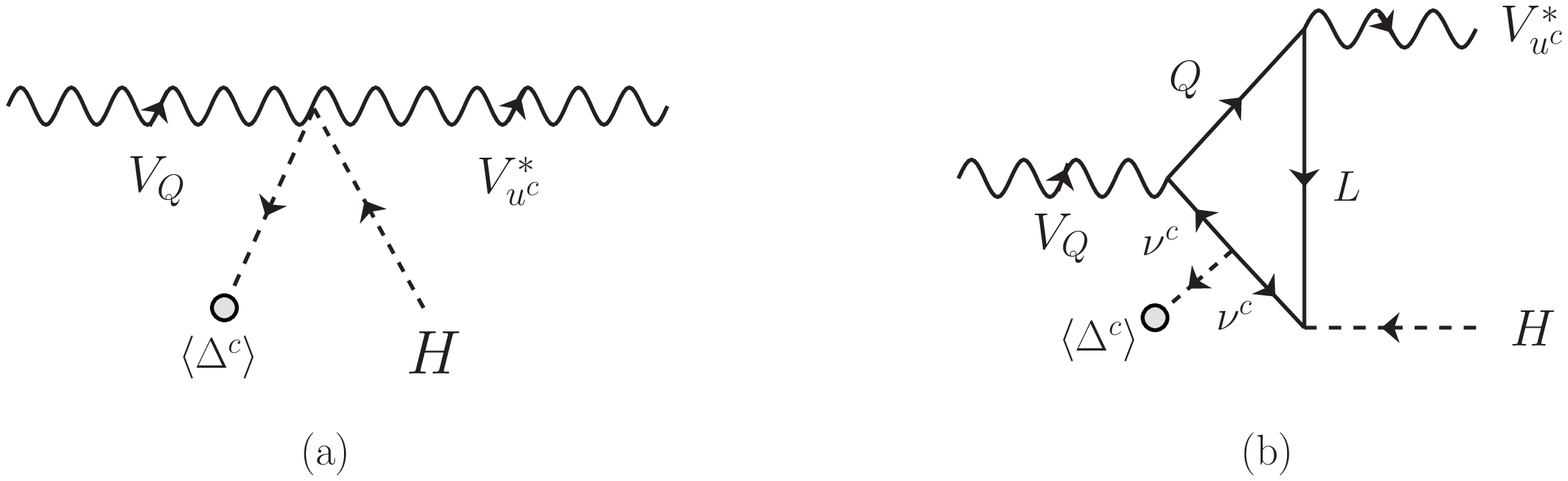}
	\caption{Tree graph and one--loop  graph responsible for $(B-L)$ asymmetry in the decay of vector gauge boson $V_Q$.}
	\label{baryo3}
\end{figure}

The $d=7$ baryon number violating operators of Eq. (\ref{dim7}) can arise by integrating out the $V_Q(3,2,1/6)$
and $V_{u^c}(\overline{3},1,-2/3)$ gauge bosons of $SO(10)$ (see Fig. \ref{gauge}), which lie outside of the $SU(5)$ subgroup.
The decay $V_Q \rightarrow V_{u^c}^* H^*$ and the conjugate decay $V_{Q}^* \rightarrow V_{u^c} H$ can produce
a primordial $(B-L)$ asymmetry at the GUT scale, which would survive down to low temperatures without being
washed out by the sphaleron interactions.  The tree--level decay diagram and the one--loop correction are
shown in Fig. \ref{baryo3}.  The vector gauge boson $V_Q$ has two--fermion decays into the following channels:
$V_{Q}^{2/3} \rightarrow \overline{u^c} \nu,\, \overline{Q} d^c,\,\overline{\nu^c} u$ for the charge $2/3$ component,
and $V_Q^{-1/3} \rightarrow \overline{u} d^c,\,\overline{u^c} e,\,\overline{\nu^c} d$ for the charge $-1/3$
component.  These decays conserve $(B-L)$, as can be seen by assigning $(B-L)(V_Q) = -2/3$.  The gauge boson
$V_{u^c}^*$ has the fermionic decays $V_{u^c}^* \rightarrow \overline{e} d,\,\overline{d^c} e^c,\,\overline{\nu} u,\,
\overline{u^c} \nu^c$, suggesting that $(B-L)(V_{u^c}^*) = 4/3$.  The decay $V_Q \rightarrow V_{u^c}^* H^*$ would then
change $(B-L)$ by $+2$, as in the case of the scalar decay $\omega \rightarrow \rho H^*$.  The $(B-L)$ asymmetry
arising from Fig. \ref{baryo3} is found to be
\begin{equation}
\epsilon_{B-L}^{(V)} =  -\frac{1}{2\pi}{\rm Im}\left[{\rm Tr}\{M_{\nu^c}\, F_4(M_{\nu^c}) \, Y_{\nu^c L H}\}\frac{g}{c^*M_{V_{u^c}}}\right]{\rm Br},
\label{asymV}
\end{equation}
with the function $F_4$ defined as
\begin{equation}
F_4(M_j) = {\rm ln}\left(1+ \frac{M_{V_{u^c}}^2}{M_j^2}\right) + \Theta\left(1- \frac{M_j^2}{M_{V_Q}^2}\right)\,\left(1- \frac{M_j^2}{M_{V_Q}^2}\right).
\end{equation}
Here we have defined the Lagrangian coefficient of the $V_Q V_{u^c} H$ vertex to be $g\, c\, M_{V_{u^c}}$.  This is a consistent
definition, since the lighter gauge boson $V_{u^c}$ would acquire a mass of order $g v_R$.  Here $c$ is a Clebsch factor of order unity,
with its value depending on the Higgs representation used for rank reduction ($126_H$ or $16_H$).

The factor ${\rm Br}$ that appears in Eq. (\ref{asymV}) is the branching ratio ${\rm Br}(V_Q \rightarrow V_{u^c}^* H^*)$.
It is determined in terms of the two partial widths as ${\rm Br} = \Gamma_1/(\Gamma_1+\Gamma_2)$, where
\begin{eqnarray}
\Gamma_1(V_Q \rightarrow V_{u^c}^* H^*) &=& \frac{g^2|c|^2}{192\pi}M_{V_Q}\left(
1-\frac{M_{V_{u^c}}^2}{M_{V_Q}^2}\right)\left\{\left(
1-\frac{M_{V_{u^c}}^2}{M_{V_Q}^2}\right)^2+ 12 \frac{M_{V_{u^c}}^2}{M_{V_Q}^2}   \right\}, \nonumber \\
\Gamma_2(V_Q \rightarrow ff) &=& \frac{g^2}{12 \pi}M_{V_Q}.
\end{eqnarray}
If we choose $M_{V_Q} = 10^{16}$ GeV, $K = \Gamma_1/H \simeq 0.016$ (with $g=0.55$ and $c=1$), so that
the dilution factor is $d\simeq 1.0$.  There is a modest suppression in $\epsilon_{B-L}^{(V)}$ arising from the branching ratio,
since ${\rm Br} \simeq 0.06$ with this choice.

\subsection{Baryon asymmetry in a class of minimal $SO(10)$ models}

We now show how the GUT scale induced asymmetry in $\omega \rightarrow \rho H^*$ decay
can consistently explain the observed value of $Y_B = (8.75 \pm 0.23) \times 10^{-11}$,
in a class of minimal $SO(10)$ models. In this class of models, a single $10_H$ and a single
$\overline{126}_H$ couple to fermions, as in  Eqs. (\ref{Yuk10})-(\ref{Yuk126}).
In SUSY models this is automatic with a single $10_H$ and $\overline{126}_H$ employed.
In non--SUSY models the $10_H^*$ can also couple ($10_H$ must be complexified to generate
realistic fermion masses), however if a Peccei--Quinn symmetry is assumed, the $10_H^*$
coupling would be absent.  
This class of models is highly constrained due to the small number of parameters
that describe the fermion masses and mixings, and leads to predictions for the neutrino oscillation parameters \cite{babu}.
In addition to generating large mixing angles for solar neutrino oscillations and for atmospheric
neutrino oscillations, these models predict a relatively large value of $\theta_{13}$, viz.,
$\sin^2 2\theta_{13} \approx (0.085-0.095)$, both in the non--supersymmetric and the supersymmetric
versions, consistent with recent results from Daya Bay and
other experiments \cite{theta13}.

To illustrate how realistic choice of parameters generate acceptable $Y_B$, we choose the
$\omega$ field to be almost entirely in the $10_H$.  We also choose $\lambda'v_R$ that appears
in Fig. \ref{baryo1} (d) to be small, so that the leading contribution to $\epsilon_{B-L}$ is
from Fig. \ref{baryo1} (c), as given in Eq. (\ref{asymc}).  In this limit, we find
\begin{equation}
\epsilon_{B-L} \approx \frac{2 \sqrt{3}}{\pi} \frac{|h_{33}f_3|^2}{|\lambda|}\left\{1+{\rm ln}\left(1+ \frac{M_\rho^2}{M_{\nu_3^c}^2}
\right)\right\} \sin\phi~.
\end{equation}
Here we have kept only the third family Yukawa couplings, which is the leading contribution, and we
have defined $\phi = {\rm arg}\{h_{33}^2 f_3^2 \lambda + \frac{\pi}{2}\}$.  Choosing $h_{33} \simeq 0.6$ (the top quark Yukawa
coupling at the GUT scale), and $\lambda = 0.25$, $v_R = 10^{16}$ GeV, $f_3 = 10^{-2}$ (so that $f_3 v_R = 10^{14}$ GeV,
consistent with the light $\nu_\tau$ mass arising via the seesaw mechanism), $\phi = 0.12$, we find $\epsilon_{B-L} = 1.6 \times 1.9 \times 10^{-5}$.
If  $M_\omega = 10^{15}$ GeV, then ${\rm Br} = 0.96$, $K=197$ so that the dilution factor is $d=
5.6 \times 10^{-4}$.  This results in a net $Y_B= 8.2 \times 10^{-11}$, consistent with observations.

While natural choices of parameters can generate acceptable $Y_B$, due to the high sensitivity of dilution factor on the
masses of the heavy particles, precise predictions are difficult to make.  For $V_Q \rightarrow V_{u^c}^* H^*$ decay
we find the process to be typically out of equilibrium so that $d\simeq 1$ for $M_{V_Q} \sim 10^{16}$ GeV.  Natural
values of the asymmetry parameter in this case is $\epsilon_{B-L} \approx 10^{-4}$.  Some dilution effects from the
$\nu^c$ interactions would be welcome in this case.

It should be mentioned that the $d=7$ operators of Eq. (\ref{lep}) also arise naturally in $SO(10)$ models, as already
noted.  The $\eta^* \rho H$ and the $\overline{\Phi} \rho H^*$ vertices arising from the $126_H$ couplings
can be used for GUT scale $(B-L)$--genesis without generating  $d=7$ nucleon decay operators.  The decays $\eta \rightarrow
\rho H$ has already been analyzed, but if these particles arise from $126_H$ they do not lead to $(B-L)$--violating
nucleon decay.

\section{Conclusion}

In conclusion, we have pointed out that the complete set of $d=7$ baryon number violating operators that lead to the selection rule $\Delta (B-L)=\pm 2$ in nucleon decay can emerge as effective low energy operators in $SO(10)$ unified theories with either a  $126_H$ or a  $16_H$ Higgs field
used for breaking the  $B-L$ gauge symmetry. The strength of these operators is unobservable in single--step models
where $SO(10)$ breaks directly down to the standard model.  In non--supersymmetric $SO(10)$ models, an intermediate symmetry is required
in order for the gauge couplings to unify correctly.  We have shown that in several instances with such an intermediate scale, the $d=7$
baryon number violating operators can lead to observable nucleon decay rates.  The decay modes are distinct from the conventional GUT--motivated
modes, and include $n \rightarrow e^-K^+,\, e^- \pi^+$, etc.  We have also identified supersymmetric scenarios where such modes may be
within reach of experiments, consistent with gauge coupling unification.

A second major result of this paper is a new way of generating $(B-L)$ asymmetry in the early universe by the decay of GUT mass particles.
It is these particles which also induce the $d=7$ nucleon decay operators.  Such an asymmetry is  sphaleron--proof, in that it does
not get erased by the effective interactions of the electroweak sphalerons.  We present several examples where consistent asymmetry
can be generated with the GUT scale decays of particles obeying  the $\Delta (B-L)= \pm 2$ selection rule.  Further, we show that in minimal $SO(10)$
models which explain the large neutrino mixing angles and predict relative large value for $\theta_{13}$, consistent with recent
experimental results, that the induced baryon asymmetry via the proposed GUT--scale mechanism is compatible with observations.
There is thus a strong connection
between neutrino oscillation parameters and baryon asymmetry in this class of models.

\section*{Acknowledgement}

The work of KSB is supported in part the US Department of Energy, Grant Numbers DE-FG02-04ER41306 and that of RNM  is supported
in part by the National Science Foundation Grant Number PHY-0968854.  KSB acknowledges helpful discussions with J. Julio.

\end{document}